\documentclass[aps,a4paper,showpacs]{revtex4}
\usepackage[colorlinks=true, pdfstartview=FitV, linkcolor=blue, citecolor=red, urlcolor=magenta,breaklinks=true]{hyperref}
\usepackage{graphicx}
\usepackage[all]{xy}
\usepackage{amsmath}
\usepackage{amssymb}
\usepackage{color}
\usepackage{subeqnarray}
\usepackage{subfigure}
\usepackage{epstopdf}

\newcommand{\bb}{\bibitem}
\newcommand{\bes}{\begin{subequations}}
\newcommand{\ees}{\end{subequations}}
\def\ben{\begin{eqnarray}}
\def\een{\end{eqnarray}}
\newcommand{\bens}{\begin{subeqnarray}}
\newcommand{\eens}{\end{subeqnarray}}
\def\be{\begin{equation}}
\def\ee{\end{equation}}

\def\o{\text{o}}
\def\tanh{\text{tanh}}
\def\sech{\text{sech}}

\def\cos{\text{cos}}
\def\arccot{\text{arccot}}

\begin{document}
\title{Kinklike structures in models of the Dirac-Born-Infeld type} 
\author{D. Bazeia} 
\author{Elisama E. M. Lima}
\author{L. Losano}  
\affiliation{Departamento de F\'\i sica Universidade Federal da Para\'iba, 58051-900 Jo\~ao Pessoa PB, Brazil}
\date{\today}

\begin{abstract}
The present work investigates several models of a single real scalar field, engendering 
kinetic term of the Dirac-Born-Infeld type. Such theories introduce nonlinearities to the kinetic part of the Lagrangian, which presents a square root restricting the field evolution and including additional powers in derivatives of the scalar field, controlled by a real parameter. In order to obtain topological solutions analytically, we propose a first-order framework that simplifies the equation of motion ensuring solutions that are linearly stable. This is implemented using the deformation method, and we introduce examples presenting two categories of potentials, one having polynomial interactions and the other with nonpolynomial interactions. We also explore how the Dirac-Born-Infeld kinetic term affects the properties of the solutions. In particular, we note that the kinklike solutions are similar to the ones obtained through models with standard kinetic term and canonical potential, but their energy densities and stability potentials vary according to the parameter introduced to control the new models. 
\end{abstract}
\maketitle


\section{Introduction}

Topological defects are localized structures that appear in several branches of physics. They are of interest in particular in high energy physics \cite{r1,DM}, where they may have implications in the evolution of the Universe during phase transitions and in other scenarios, and also in condensed matter systems \cite{CMab,CMa}, where they may appear to model interfaces intersecting distinct regions of the space. Among the diversity of defects that arise in classical field theories, the one-dimensional static solutions of equations of motion described by scalar fields, known as kinks, are the simplest topological structures, which have been studied by several authors; see, e.g.,
Refs.~\cite{r1,DM,CMab,CMa,dutra,Romanczukiewicz,scater,Randall,brane,thickbrane,compactbrane,kfield}. For instance, kink structures may be employed in the study of kink-antikink pairs production from particle scattering \cite{dutra} and in investigations involving kink-antikink collisions \cite{Romanczukiewicz,scater}. They are also important to researches based on  braneworld scenarios, accompanied by the thick or compact profile, with a single extra dimension of infinite extent \cite{Randall,brane,thickbrane,compactbrane}. Besides, they have been studied within the context of theories with modified kinematics with other motivations \cite{kfield}.

Scalar field theories with nonstandard kinetic term admit the presence of  topological structures, also known as $k$-defects, that lead to interesting consequences for several physical situations \cite{Babichev1,Babichev2,SPGD,SPGD2,FOFGD,Adam,Rosenau,DMGD,Olechowski,BGD}. For instance, in Refs.~\cite{Babichev1,Babichev2}, general topological properties of $k$-defects and $k$-vortices, where the gauge field is preserved in its canonical form, were discussed in connection with cosmological applications. In this case in particular, the nonlinearities there introduced by the kinetic term lead to  the presence of a new mass scale which might alter some characteristics  associated to the formation of cosmic strings during phase transitions in the early universe. In Refs.~\cite{SPGD,SPGD2} the authors develop a study containing the linear stability of defect structures with modified profile, including the presence of compactons \cite{Adam,FOFGD,DMGD}, that is, of solutions with compact support \cite{Rosenau}; similar modifications are also analyzed in the braneworld contexts in Refs.~\cite{Olechowski,BGD}. 

Another interesting issue arises when the kinetic term is specifically of the Dirac-Born-Infeld (DBI) type \cite{Sarangi,Babichev2,VortexBI,Brown,Andrews,Garcia}. Originally, the DBI theory was introduced to eliminate problems with divergence of electron self-energy  in classical electrodynamics \cite{DBI}. Posteriorly, such reasoning passed to be explored in studies involving fields with unusual kinetic terms; for example, it has been used for the construction of instanton solutions \cite{Brown}, for the considerations of global strings \cite{Sarangi}, for the description of global vortex solutions \cite{Babichev2,VortexBI}, for the representation of twinlike models which develop through distinct systems supporting exactly the same topological structure \cite{Andrews}, and also for the formulation of models admitting soliton solutions \cite{Garcia}. Following these lines, the present paper consider a modification on the field dynamics, where the kinetic part is DBI-like, and we study the existence of kinklike defects for several new models not yet investigated in the literature. Generally, these kinds of modifications add new nonlinear terms to the Lagrangian and make the study very complicate to be solved exactly; nevertheless, the aim of the current work is to solve analytically the equations of motion ensuring the presence of defect structures and discuss some of their features, such as the energy density and stability potential. 

To achieve these goals we establish a first-order framework that simplifies the equations of motion, 
in a way compatible with the description given by Bogomoln’yi-Prasad-Sommerfeld (BPS) \cite{BPS}. We adopt the deformation procedure \cite{deform} developed to help us to search for exact solutions in systems with generalized dynamics \cite{DMGD}. This method has been successful in providing results concerning the presence of analytical solutions for physical problems engendering nonlinear dispersion \cite{DMGD} and so it will be useful here. In this work we deal with two categories of scalar field models, one described by polynomial self-interactions and the other by nonpolynomial ones. 

In order to explore how the DBI kinetic effects modify the properties of the topological solutions, we organize 
the work as follows. In the next section, we review the general formalism for
a field theory with nonstandard kinetic term. In Sec.~\ref{sec-2} we specify the field dynamics 
consistently with the DBI concept. We suggest a first-order treatment which solves the equation of motion and ensures the linear stability of the solutions. Also, in order to explore further potentials we incorporate the deformation procedure. In Sec. \ref{sec-3} we present several examples of DBI models,  engendering polynomial and nonpolynomial interactions. There, all the solutions are found exactly and their main characteristics are studied in details. We also note that, if the parameter which regulates the high-order powers in the derivatives of $\phi$ is very large, then the outcome can be compared with the corresponding standard theory. We end the work including our comments and conclusions in Sec.~\ref{sec-com}.

\section{GENERALIZED FORMALISM}
\label{sec-1}

We start dealing with a generalized theory of a real scalar field, described by the Lagrangian density
\be
\label{lagran0}
{\cal L}(\phi,X)=F(X)-V(\phi),
\ee
where
\be
X = \frac{1}{2}\partial_{\mu}\phi\partial^{\mu}\phi,
\ee
and therefore we identify $F(X) = X$ to represent the standard model. 

The system is defined in $(1, 1)$ spacetime dimensions, so $\phi = \phi(x, t)$, and we use the metric $(+,-)$. For simplicity, we consider dimensionless field, spacetime coordinates and coupling constants. The above Lagrangian density allows us to obtain the energy-momentum tensor
\be
T_{\mu\nu}=F_{X}\partial_{\mu}\phi\,\partial_{\nu}\phi-\eta_{\mu\nu}\,{\cal L}.
\ee
Here, $F_{X} = dF/dX$ and the equation of motion for the scalar field $\phi(x, t)$ has the form
\be
\label{eom0}
\partial_{\mu}(F_{X}\partial^{\mu}\phi)+V_{\phi}=0,
\ee
where $V_\phi=dV/d\phi$. Since we are searching for defect structures, we consider static configurations, $\phi = \phi(x)$. The equation of motion becomes
\be
\label{eom1}
(F_X+2XF_{XX})\phi''=V_{\phi},
\ee
where $F_{XX} = d^2F/dX^2$ and the prime represents derivative with respect to $x$ and now $X = -\phi'^2/2$. The equation \eqref{eom1} can be reduced to first order, giving
\be
\label{fore}
F-2XF_{X}-V=0.
\ee
Note that, the above equation obeys the stressless condition, $T_{11} = 0$. Moreover, the energy density can be written as $\rho(x) = T_{00} =-F +V$, and so from the condition \eqref{fore} we have
\be
\label{rho}
\rho(x)=F_{X}\phi'^2.
\ee
Now, following the formalism developed in Ref.~\cite{FOFGD}, we introduce a new function $W = W(\phi)$, such that
\be
\label{nfoe}
F_X \phi'=W_{\phi}.
\ee
It allows to write the energy density as $\rho(x) = W_{\phi} \phi' = dW/dx$, and so the energy becomes
\be
E=W(\phi(x\rightarrow\infty))-W(\phi(x\rightarrow-\infty)).
\ee
On the other hand, the first-order equation \eqref{fore} may also be written as
\be
\label{fpot}
W_{\phi}\phi'=-{\cal L}.
\ee
Also, deriving \eqref{nfoe} with respect to $x$, the equation of motion \eqref{eom1} is satisfied whether
\be
\label{ssoe}
V_{\phi}=W_{\phi\phi}\phi',
\ee
that is also a first-order differential equation which combined with \eqref{nfoe} provides solutions for the equation of motion.

Let us now focus on the linear stability. In order to implement this, we introduce a small perturbation around the static solution, $\phi(x, t) = \phi(x) + \eta(x)\cos(\omega t)$, which is substituted in \eqref{eom0} by taking into account the fluctuations up to first order. The procedure is similar to the one studied before in \cite{SPGD}. Making the following changes for new variables
\be
dx=A dz \:\:\: \:\:\: \mbox{and} \:\:\: \:\:\: \eta=\frac{u}{\sqrt{F_{X}A}},
\ee 
where
\be
A^2=\frac{2XF_{XX}+F_X}{F_X},
\ee
one gets a Schr\"odinger-like equation, $Hu(z) = \omega^2u(z)$, such that
\be
\label{eshor}
\left(-\frac{d^2}{dz^2}+U(z)\right)u=\omega^2u,
\ee
and
\be
\label{spz}
U(z)=\frac{(\sqrt{F_{X}A})_{zz}}{\sqrt{F_{X}A}}+\frac{V_{\phi\phi}}{F_X}.
\ee
The linear stability depends upon eigenvalues $\omega^2$ nonnegative, which in turn depend on the shape of the potential $U(z)$ for each specific problem under investigation. An interesting issue here is the fact that the first-order formalism established above is characterized by the appearance
of a function $W(\phi)$ which can help us to factorize the Schr$\ddot{\o}$dinger-like equation \eqref{eshor} to ensure that $\omega$ is a real quantity, with $\omega^2$ nonnegative. Moreover, there is the possibility of $\omega$ be zero, providing the zero mode as the corresponding eigenstate. All of these properties will be explored in the next sections, for a single scalar field with Dirac-Born-Infeld kinetic term \cite{Andrews,Garcia}.

\section{DIRAC-BORN-INFELD DYNAMICS}
\label{sec-2}

Here we adopt a field theory in which the kinetic term is of the Dirac-Born-Infeld type \cite{DBI}, having the form
\be
\label{FBI}
F(X)=-a^2\left(\sqrt{1-\frac{2X}{a^2}}-1\right),
\ee
where $a$ is a dimensionless real parameter, which controls the high-order powers in the covariant derivative of the scalar field. It is easy to see that by expanding the equation above for $1/a^2 << 1$, one gets back to the canonical kinetic term, that is,
\be
F(X)=X+O\left(\frac{1}{a^2}\right).
\ee
Based on the first-order description presented in Sec.~\ref{sec-1}, we assume static configurations so that the equation \eqref{nfoe} becomes
\be
\label{fow}
\phi'=\frac{W_{\phi}}{\sqrt{1-\dfrac{W_{\phi}^2}{a^2}}},
\ee
and replacing it in \eqref{fpot} we obtain a specific form for $V(\phi)$,
\be
\label{potw}
V(\phi)=-a^2\left(\sqrt{1-\dfrac{W_{\phi}^2}{a^2}}-1\right).
\ee
In addition, there is another equation related to Eq.~\eqref{fow} obtained by taking $x\rightarrow -x$. Also, one can check through \eqref{fow} and \eqref{potw} that the relation \eqref{ssoe} is satisfied;  thus the equation of motion of this system is obeyed through the treatment described above,  see Eq.~\eqref{eom1}; and hence we are interested in constructing $W$ functions such that the first-order equations support analytical solutions. In this case, the energy density  \eqref{rho} is given by
\be
\label{rhofo}
\rho(x)=\frac{\phi'^2}{\sqrt{1+\dfrac{\phi'^2}{a^2}}}=\frac{W_{\phi}^2}{\sqrt{1-\dfrac{W_{\phi}^2}{a^2}}}.
\ee

We emphasize that under the circumstances presented above, the Lagrangian \eqref{lagran0} is explicitly given by
\be\label{lagra}
{\cal L}(\phi,X)= -a^2\left(\sqrt{1-\frac{2X}{a^2}}-\sqrt{1-\dfrac{W_{\phi}^2}{a^2}}\right),
\ee
and hence when $1/a^2<<1$, we get to the standard formalism  which develops when the scalar field acquires usual dynamics
\be
{\cal L}(\phi,X)=X-\frac{1}{2}W_{\phi}^2 + O\left(\frac{1}{a^2}\right),
\ee
in such a way that $V(\phi)\approx \frac{1}{2}W_{\phi}^2$, the first-order equation for the static field becomes $\phi'\approx W_\phi$, and the energy density assume the form $\rho(x)\approx \phi'^2 \approx W_\phi^2$; thereby ensuring an approach compatible with the description for BPS states \cite{BPS}. The form of the Lagrangian \eqref{lagra} ensures the first order formalism when the scalar field engenders the DBI dynamics.  

\subsection{Linear stability}

As seen previously, the study of linear stability involves the change of variables for $dx=A dz$, such that
\be
\label{fz}
\phi'=\frac{\phi_z}{\sqrt{1-\dfrac{\phi_z^2}{a^2}}}.
\ee
Consequently the Eq.~\eqref{spz} becomes 
\be
U(z)=\left(\left(\sqrt{1-\dfrac{\phi_z^2}{a^2}}\right)_{zz}+V_{\phi\phi}\right)\left/ \sqrt{1-\dfrac{\phi_z^2}{a^2}}\right. .
\ee
Using \eqref{fow} and \eqref{fz} we can write: $\phi_z=W_{\phi}$, and this leads to 
\be
\label{spotg}
U(z)=W_{\phi\phi}^2+W_{\phi}W_{\phi\phi\phi}.
\ee
It permits to describe the Schr$\ddot{\o}$dinger-like equation \eqref{eshor} in a factorizable manner, ${\bf A}^{\dagger}{\bf A}\, u=\omega^2u$, where
\be
{\bf A}=-\frac{d}{dz}+W_{\phi\phi}  \:\:\: \:\:\: \mbox{and}   \:\:\: \:\:\:  {\bf A}^{\dagger}=\frac{d}{dz}+W_{\phi\phi}.
\ee
This shows that there is no bound state with negative eigenvalues and ensures the linear stability of the static solutions. Another interesting fact is that there is a zero mode associated to the null eigenvalue, which can be written as
\be
u_0(z)=c\, \phi_z= c\, W_{\phi},
\ee  
where $c$ is a normalization constant. This result can be demonstrated from the second order equation for static field, which can also be expressed as 
\be
\left(F_X \phi' \right)'=V_{\phi}.
\ee
By taking the derivative of this equation with respect to $x$, after rewriting it in terms of $z$ and $W_{\phi}$, and using \eqref{potw}, we get 
\be
\left(-\frac{d^2}{dz^2}+W_{\phi\phi}^2+W_{\phi}W_{\phi\phi\phi}\right)\phi_z=0,
\ee
which is a Schr\"odinger-like equation with zero eigenvalue, so we identify the derivative of $\phi(z)$ as the zero mode.

\subsection{Deformation procedure}

We now focus attention on new models, which can be analytically soluble for a real scalar field under the DBI perspective. Toward this goal, we use the deformation procedure extended for modified kinematics, constructed before in Ref.~ \cite{DMGD}. This method consists of adding a deforming function, $f(\chi)$, which connects a system with potential $V(\phi)$ that support analytical solutions to another one, for which one wants to find the solutions, $\tilde{V}(\chi)$.  Therefore, we introduce another system described by the scalar field $\chi(x, t)$ with Lagrange density analogous to \eqref{lagran0}, such that
\be
\label{lagran2}
{\cal L}(\chi,Y)=F(Y)-\tilde{V}(\chi),
\ee
where now $Y = \frac{1}{2}\partial_{\mu}\chi\partial^{\mu}\chi$ and the noncanonical kinetic term is 
\be
\label{FBIq}
F(Y)=-a^2\left(\sqrt{1-\frac{2Y}{a^2}}-1\right).
\ee
Also, we suppose that this new system satisfies the first-order construction that appears in Eqs.~\eqref{fow} and \eqref{potw}, that is,
\ben
\label{potwq}
\tilde{V}(\chi)&=&-a^2\left(\sqrt{1-\dfrac{W_{\chi}^2}{a^2}}-1\right), \\
\label{fowq}
\chi'&=&\frac{W_{\chi}}{\sqrt{1-\dfrac{W_{\chi}^2}{a^2}}}.
\een

Observe that the second member of Equation \eqref{fow}  can be written as a function of the field $\phi$ itself, so for simplicity, one  makes: $\phi'=R(\phi)$. Similarly one can write: $\chi'=S(\chi)$. By means of the deformation function $\phi \rightarrow f(\chi)$, we can say that  $\phi' \rightarrow f_{\chi} \chi'$, then
\be\label{SR}
S(\chi)=\frac{R(\phi \rightarrow f(\chi))}{f_{\chi}}.
\ee
Employing the Eq.~\eqref{fowq},
\be
W_{\chi}=\pm\frac{S(\chi)}{\sqrt{1+\dfrac{S(\chi)^2}{a^2}}}, 
\ee
consequently,  the so-called deformed potential can be expressed as
\be
\label{DP}
\tilde{V}(\chi)= -a^2\left(\frac{1}{\sqrt{1+\dfrac{S(\chi)^2}{a^2}}}-1\right),
\ee
and the energy density now becomes
\be
\label{Drho}
\rho(x)= \frac{W_{\chi}^2}{\sqrt{1-\dfrac{W_{\chi}^2}{a^2}}}=\frac{S(x)^2}{\sqrt{1+\dfrac{S(x)^2}{a^2}}}. 
\ee

The method implies that if we know $\phi(x)$ is solution of the model ${\cal L}(\phi,X)$, then from the inverse deformation function, $\chi(x) = f^{-1}(\phi(x))$, one obtains  the  static solution for the new model ${\cal L}(\chi,Y)$. In the next section, we exemplify this fact with several examples. 

\section{Illustrations}
\label{sec-3}

To show explicitly how the procedure works for the DBI case, we firstly propose a new potential that supports kink-like solution and is analytically soluble. In the sequence, we employ the deformation procedure to find new models, whose solutions are also found analytically. This section involve  two classes of models, described by polynomial and nonpolynomial potentials.

\subsection{Polynomial interactions}

\subsubsection{Model I}
\label{model1}

\begin{figure}
\includegraphics[width=4.2cm,height=4.2cm]{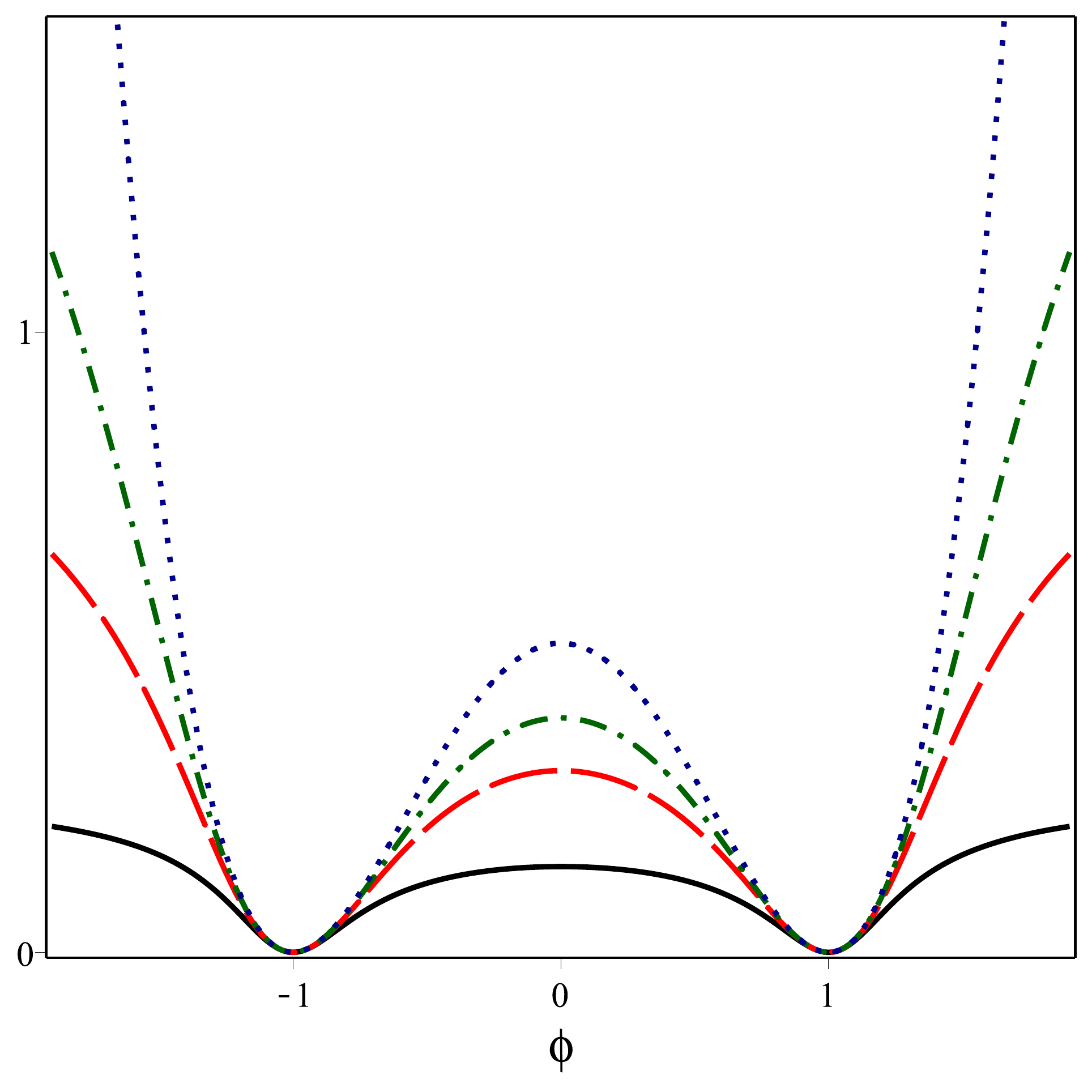}
\includegraphics[width=4.2cm,height=4.2cm]{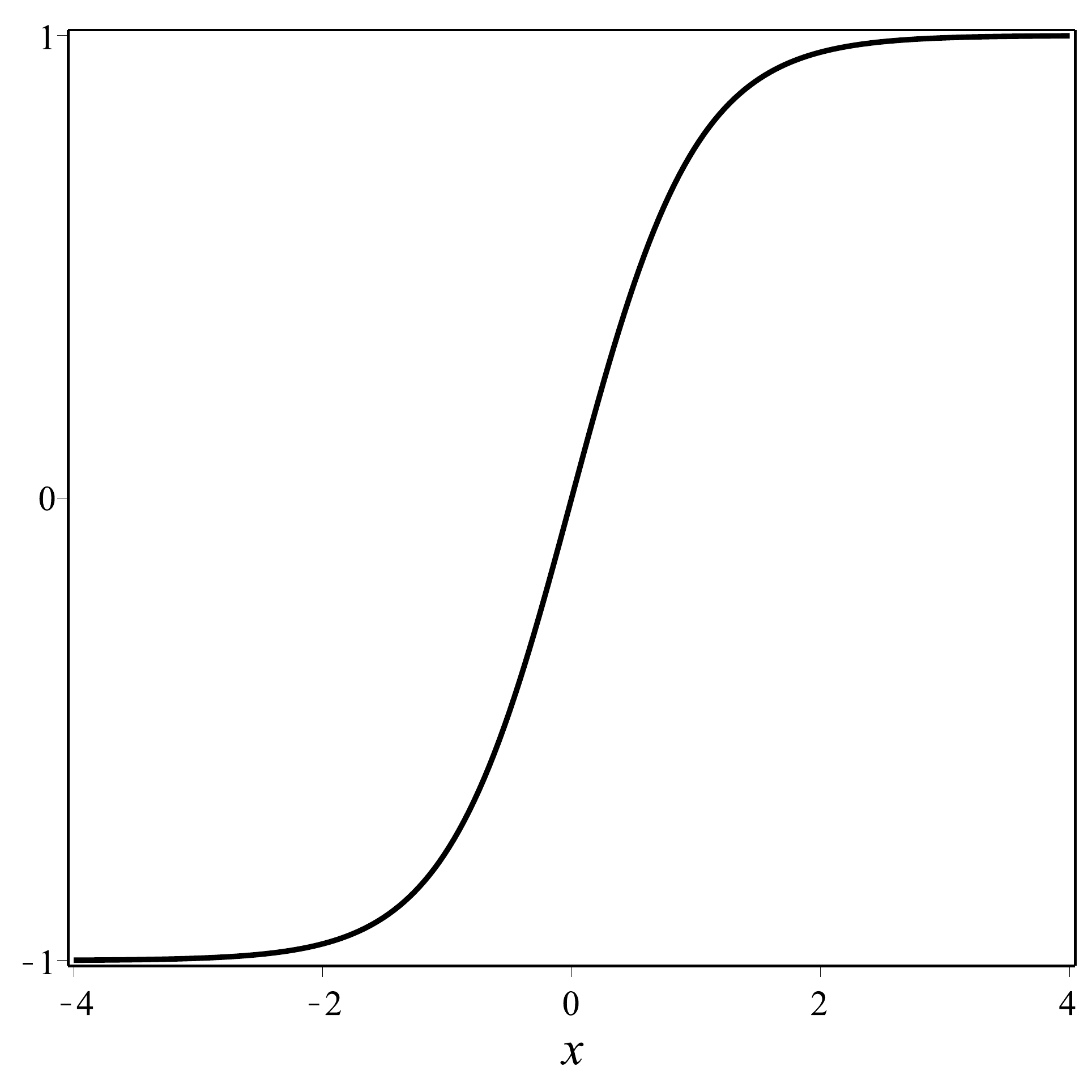}
\includegraphics[width=4.2cm,height=4.2cm]{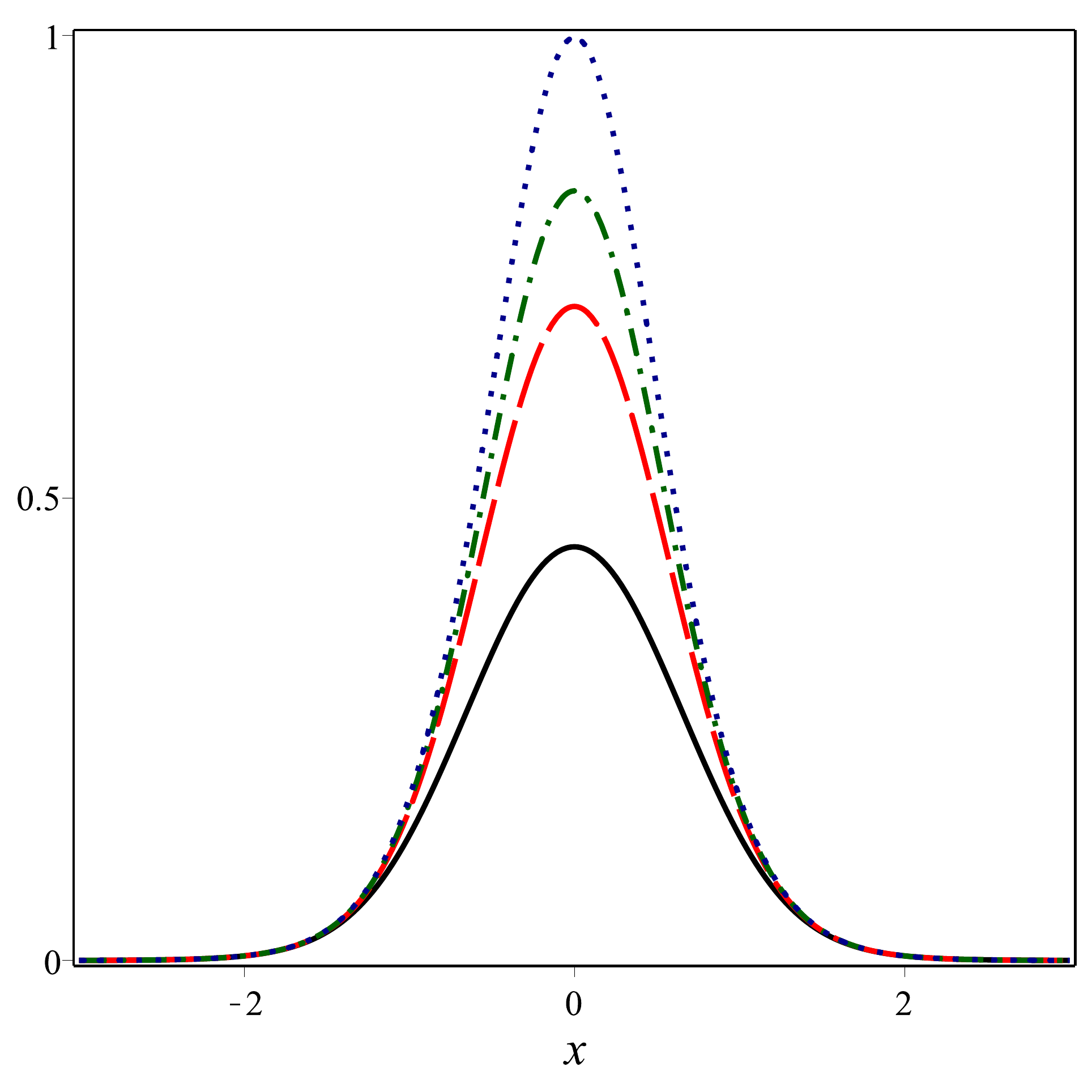}
\includegraphics[width=4.2cm,height=4.2cm]{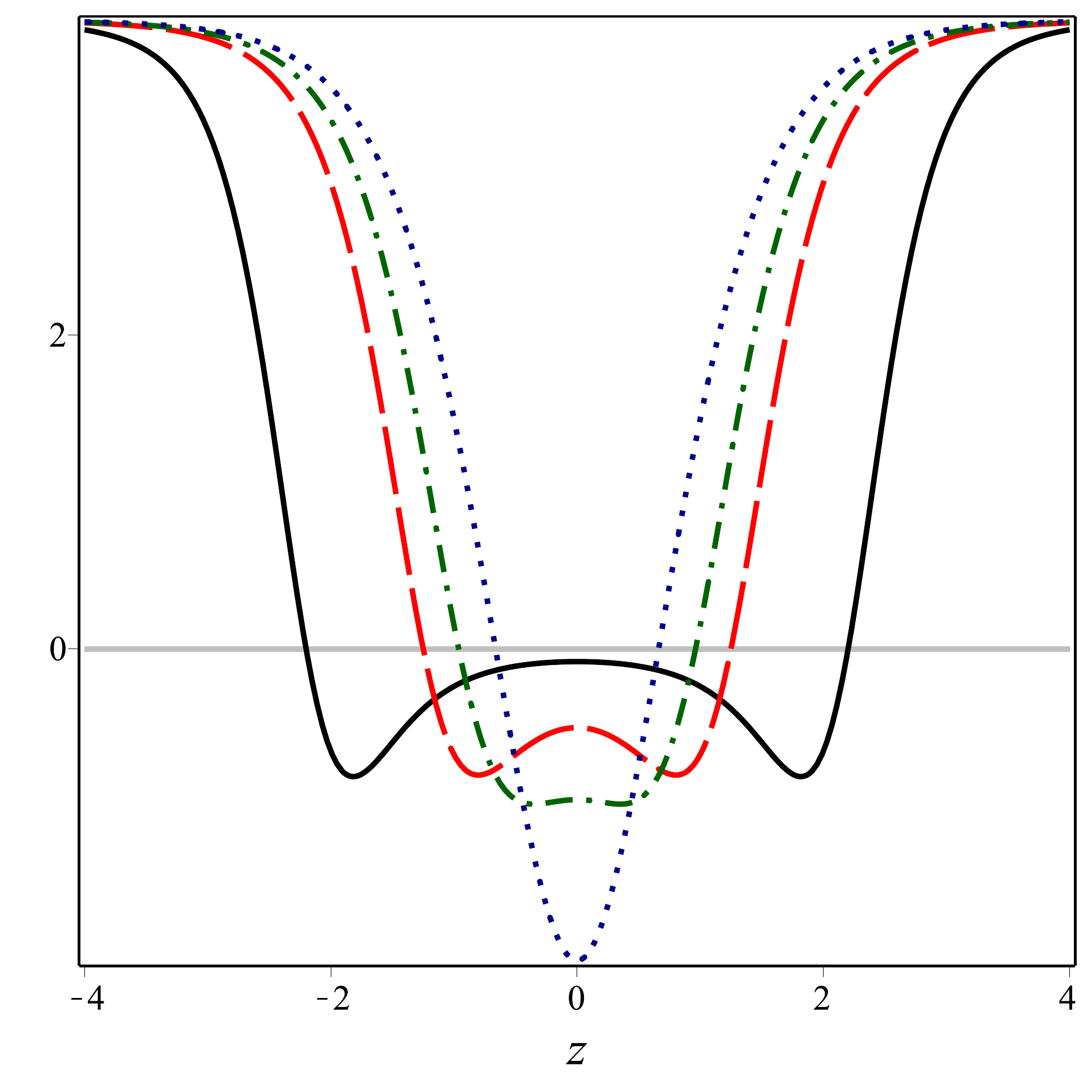}
\caption{We illustrate from left to right, the potential \eqref{pot1}, the kink solution \eqref{sol1}, the energy density \eqref{rh1} and the stability potential \eqref{spotg}, which is evaluated numerically. We are using $a=0.5,1,1.5,15$, represented by solid (black), dashed (red), dot-dashed (green) and dotted (blue) lines, respectively.}
\label{fig:1}
\end{figure}

Initially  we propose the model  with
\be
W_{\phi}=\frac{1-\phi^2}{\sqrt{1+\dfrac{1}{a^2}(1-\phi^2)^2}}.
\ee
So from Eq.~\eqref{potw}
\be
\label{pot1}
V(\phi)=-a^2\left(\frac{1}{\sqrt{1+\dfrac{1}{a^2}(1-\phi^2)^2}}-1\right).
\ee
This potential is nonnegative and has two minima at $\bar{\phi}=\pm 1$ and one maximum at the origin; see Fig.~\ref{fig:1}. The first-order equation \eqref{fow} leads to
$\phi'=1-\phi^2$, whose solution is independent of $a$, that is,
\be
\label{sol1}
\phi(x)=\pm \tanh(x),
\ee
where we have considered the possibility $x \rightarrow -x$, and we identify the top sign $(+)$ with the kink solution and the bottom sign $(-)$ with the anti-kink. Here the energy density \eqref{rhofo} gives
\be
\label{rh1}
\rho(x)=\frac{\sech^4(x)}{\sqrt{1+\dfrac{1}{a^2}\sech^4(x)}}
\ee

In Fig.~\ref{fig:1} we illustrate the potential \eqref{pot1}, the solution \eqref{sol1}, the energy density \eqref{rh1}, and the stability potential got through \eqref{spotg}, which was calculated numerically, since the change of variables $x \rightarrow z$ involves a nonanalytical integral expression. Thus we see that, $V(\phi)$ reaches the constant value $a^2$ for large $\phi$, and  the height of its maximum decreases as the parameter  $a$ decreases. The scalar field interpolates between the two distinct minima of the potential which constitute a topological sector. Also note that the stability potential, $U(z)$, has a local maximum around the origin which changes to become the minimum for $a^2\geq3$; however, such behavior is not enough to split the energy density as it happens to certain systems, as for instance in braneworld contexts with  standard or generalized gravity \cite{split}, and yet in the flat scenarios studied in Refs.~\cite{flatsplit,lima}. 

Moreover,  when $1/a^2 << 1$,  the results presented in this subsection coincide with the ones described by the standard theory under the well-known self-interaction $\phi^4$ \cite{r1}. Accordingly, $V(\phi)\approx \frac{1}{2}(1-\phi^2)^2$; the static solutions remain the same, since the solution  \eqref{sol1} independ of $a$; the energy density becomes  $\rho(x) \approx \sech^4(x)$; and the stability potential acquires the form
\be
U(x) \approx 4-6\,\sech^2(x),
\ee
that is a potential of the modified P\"oschl-Teller type \cite{Flugge}, which has two bound states, one zero mode with eigenvalue $\omega_0^2 = 0$ and one excited state with $\omega_1^2 = 3$. 

We now manipulate the potential $V(\phi)$ given above to formulate new deformed potentials. For that, one uses  $R(\phi)=1-\phi^2$, so that the equation \eqref{SR} becomes
\be
S(\chi)=\frac{1-f(\chi)^2}{f_{\chi}}.
\ee
Also note that, if we consider another deformation function gives by $1/f(\chi)$, it provides the same expression above for $S(\chi)$ and, naturally, the same deformed potential $\tilde{V}(\chi)$. This interesting property will be very important in obtaining solutions for the nonpolynomial models to be introduced below.

\subsubsection{Model II}

If we consider the deforming function $f(\chi)=2\chi^2-1$,  then
\be
S(\chi)=\chi(1-\chi^2).
\ee
And the deformed potential \eqref{DP} turns into
\be
\label{pot6}
\tilde{V}(\chi)=-a^2\left(\frac{1}{\sqrt{1+\dfrac{1}{a^2}\chi^2(1-\chi^2)^2}}-1\right),
\ee
which has three minima, $\bar{\chi}=0,\pm 1$, defining two topological sectors with solutions 
\be
\label{sol6}
\chi(x)=\pm \sqrt{\frac{1\pm\tanh(x)}{2}},
\ee
whose energy densities, see Eq.~\eqref{Drho}, are
\be
\label{rho6}
\rho(x)=\frac{1}{8}\frac{\sech^2(x)(1\mp \tanh(x))}{\left(1+\dfrac{1}{8a^2}\sech^2(x)(1\mp\tanh(x))\right)}.
\ee

\begin{figure}
\includegraphics[width=4.2cm,height=4.2cm]{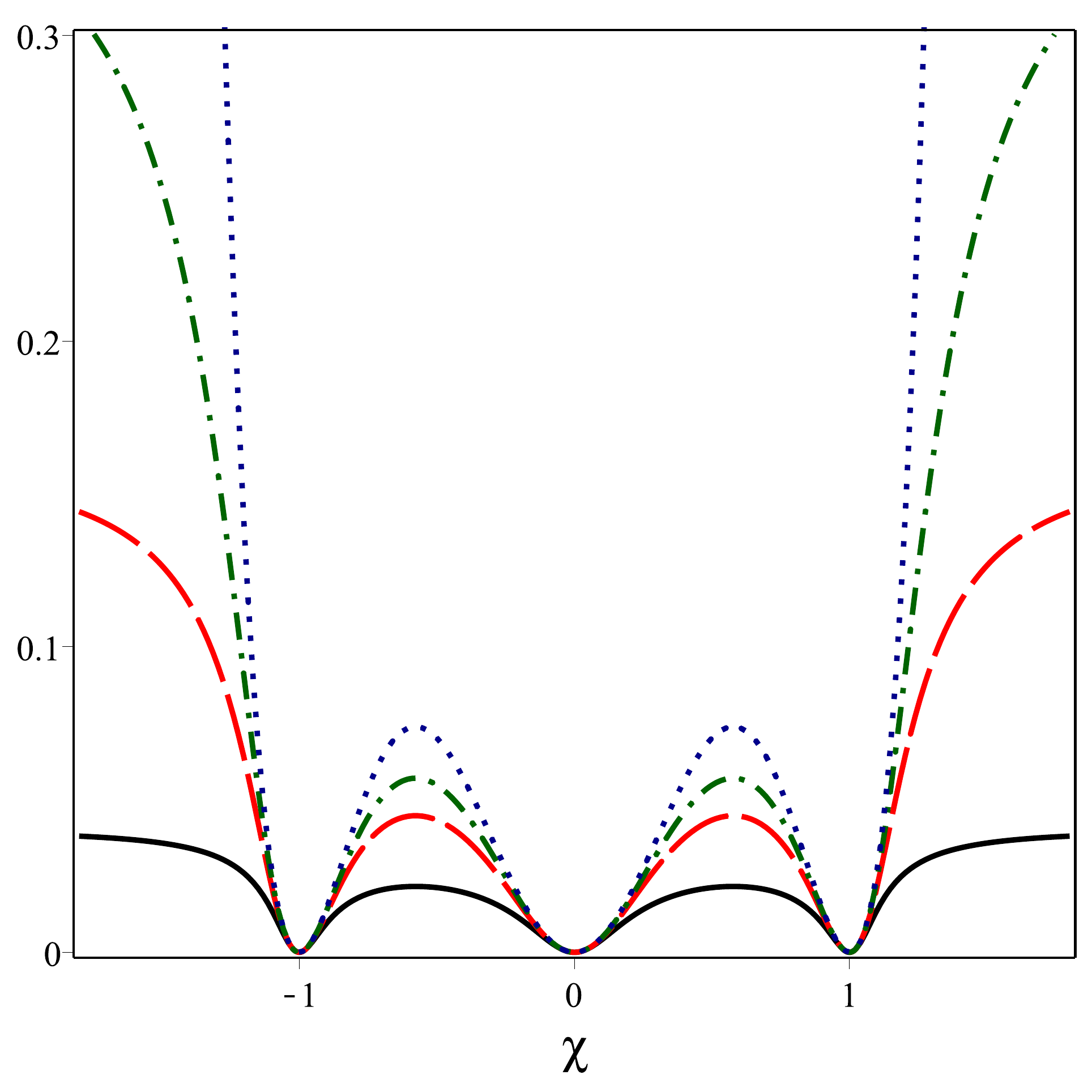}
\includegraphics[width=4.2cm,height=4.2cm]{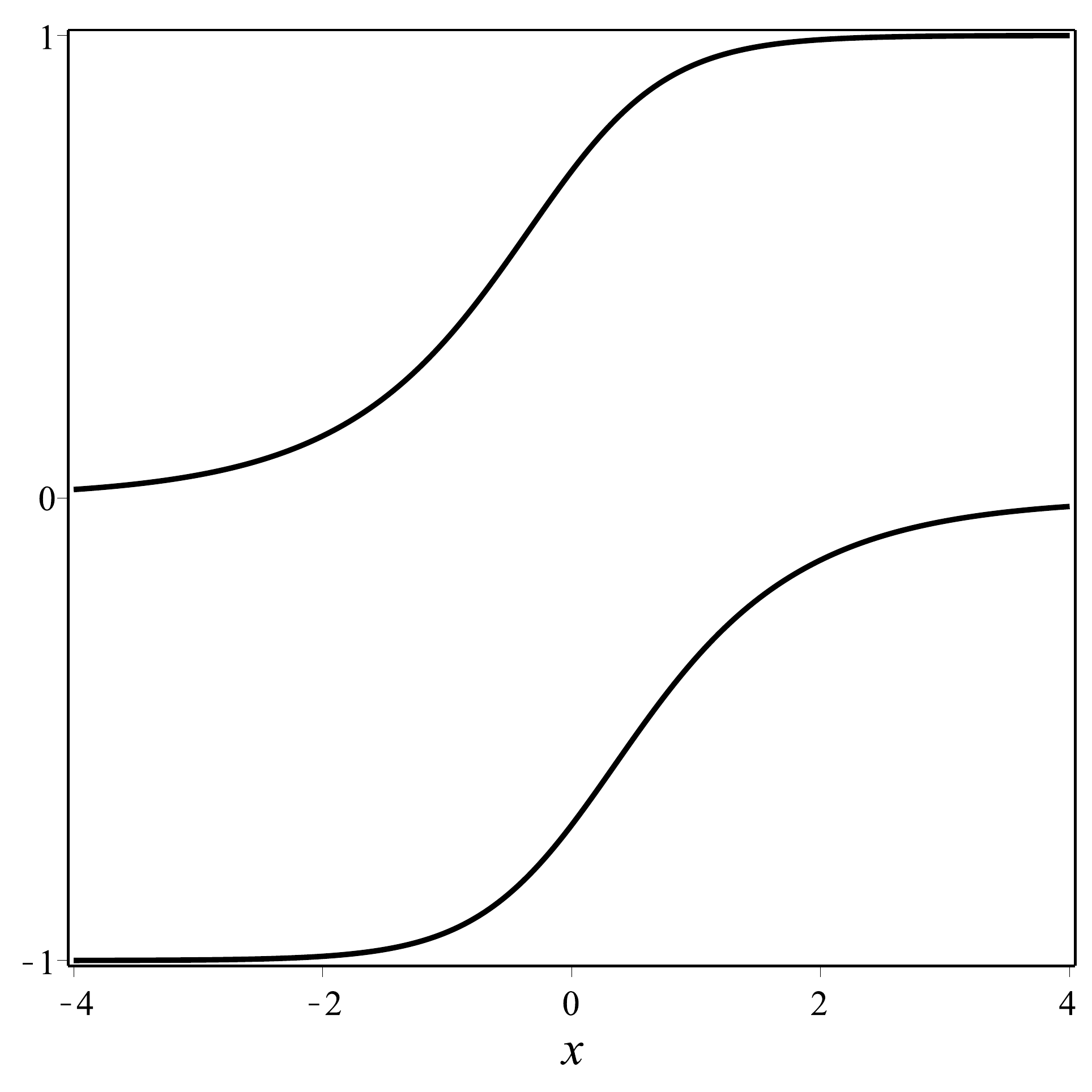}
\includegraphics[width=4.2cm,height=4.2cm]{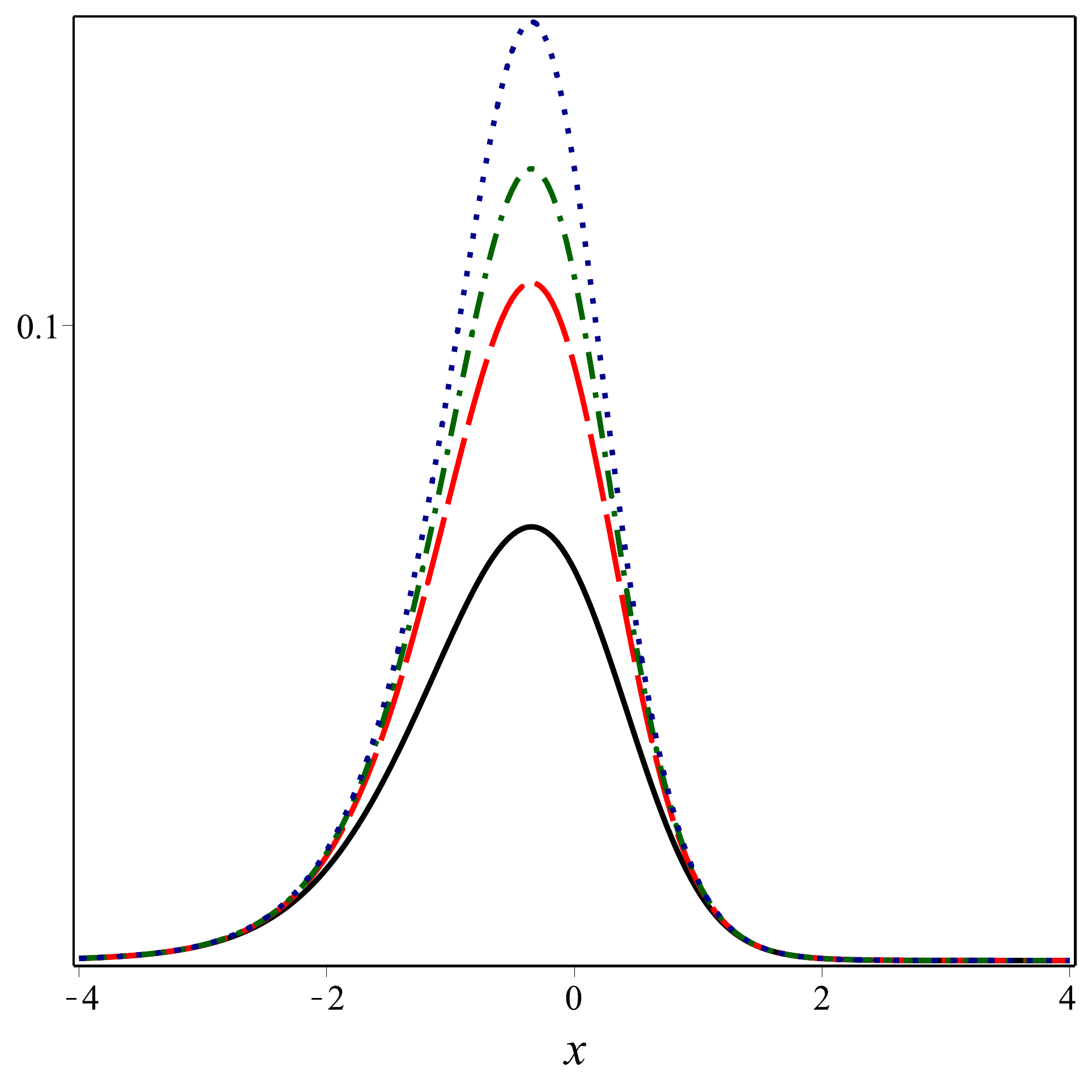}
\includegraphics[width=4.2cm,height=4.2cm]{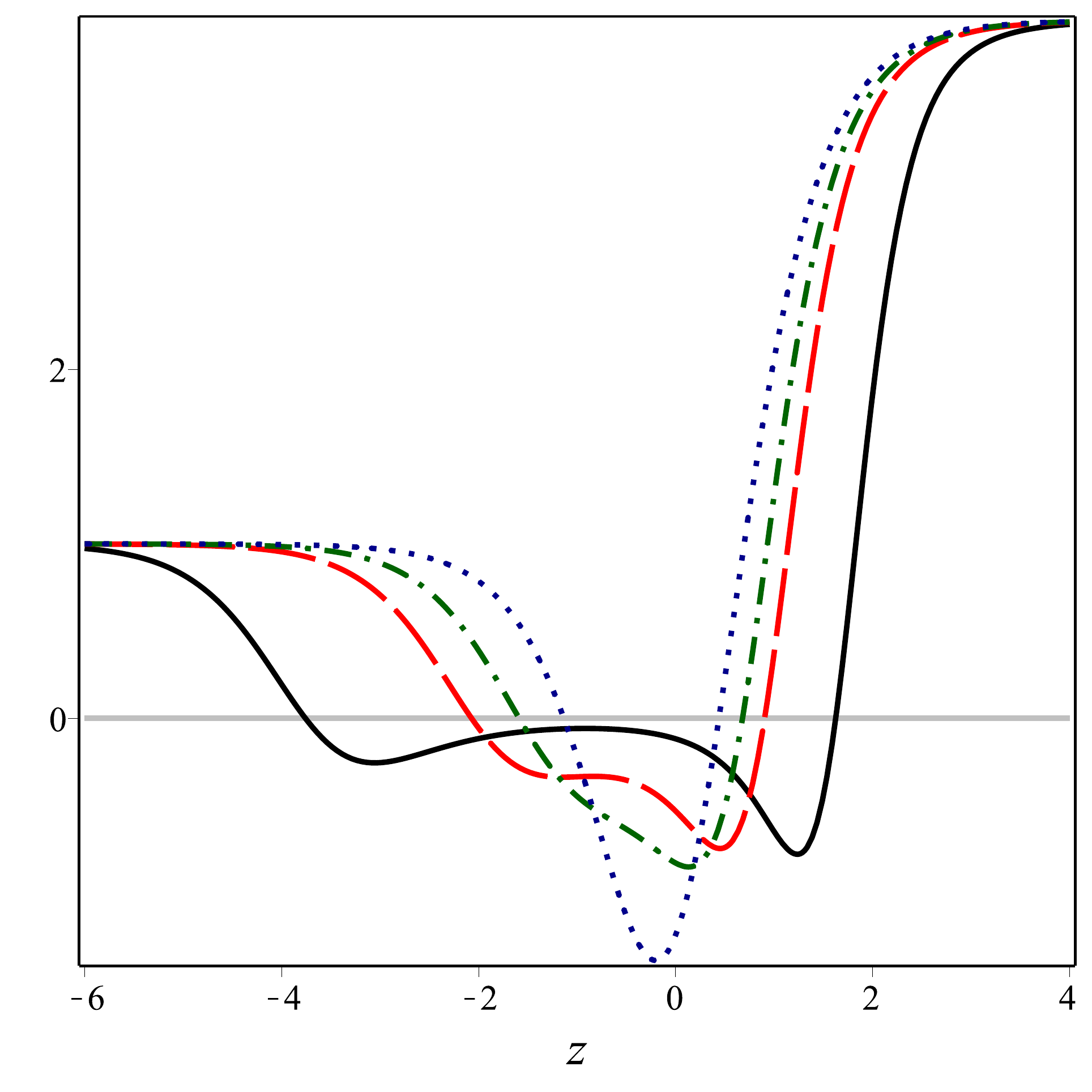}
\caption{We illustrate from left to right, the potential \eqref{pot6}, the kink solutions \eqref{sol6}, the energy density \eqref{rho6} and the stability potential, which is evaluated numerically. The solutions are associated with the kink (anti-kink) of the sector $0\rightarrow 1$ $(-1 \leftarrow 0)$. There is another similar plot, reflected, which is related to the anti-kink  (kink) of the sector $0 \leftarrow 1$ $( -1\rightarrow 0)$. Here we are using $a=0.2,0.4,0.6,6$, represented by solid (black), dashed (red), dot-dashed (green) and dotted (blue) lines, respectively.}
\label{fig:2}
\end{figure}

In Fig.~\ref{fig:2}, one depicts the potential \eqref{pot6}, which is characterized by two topological sectors; the  kink solutions related to both topological sectors of $\tilde{V}$, see Eq.~\eqref{sol6};  the energy density associated to the kink (anti-kink) of the right (left) sector, Eq.~\eqref{rho6}, since the one corresponding to the  anti-kink (kink) of the right (left) sector is equivalent to that one reflected; and the stability potential  which is obtained numerically for this system,  likewise the previous model. As can be seen, $\tilde{V}(\chi)$ also reaches the constant value $a^2$ for large $\chi$, and  the height of its maxima also decreases as $a$ decreases. The scalar field has two kink solutions interpolating between the distinct  sectors:  $0\rightarrow 1$ (right sector) and $-1\rightarrow 0$ (left sector). Here, the stability potential is asymmetric, differently from the previous model which was symmetric; furthermore, it has a peculiar behavior for small $a$; otherwise, for $1/a^2 << 1$, it  develops the form
\be
U(x) \approx \frac{5}{2} \pm \frac{3}{2}\tanh(x)-\frac{15}{4}\sech^2(x),
\ee
where there is only one bound state corresponding to $\omega^2=0$, which is the zero mode \cite{lohe}. Assuming $a$ large, the outcome got from this subsection are equivalent to the one described by the standard theory with the $\phi^6$ potential.

\subsubsection{Family I}

\begin{figure}
\includegraphics[width=4.4cm,height=4.4cm]{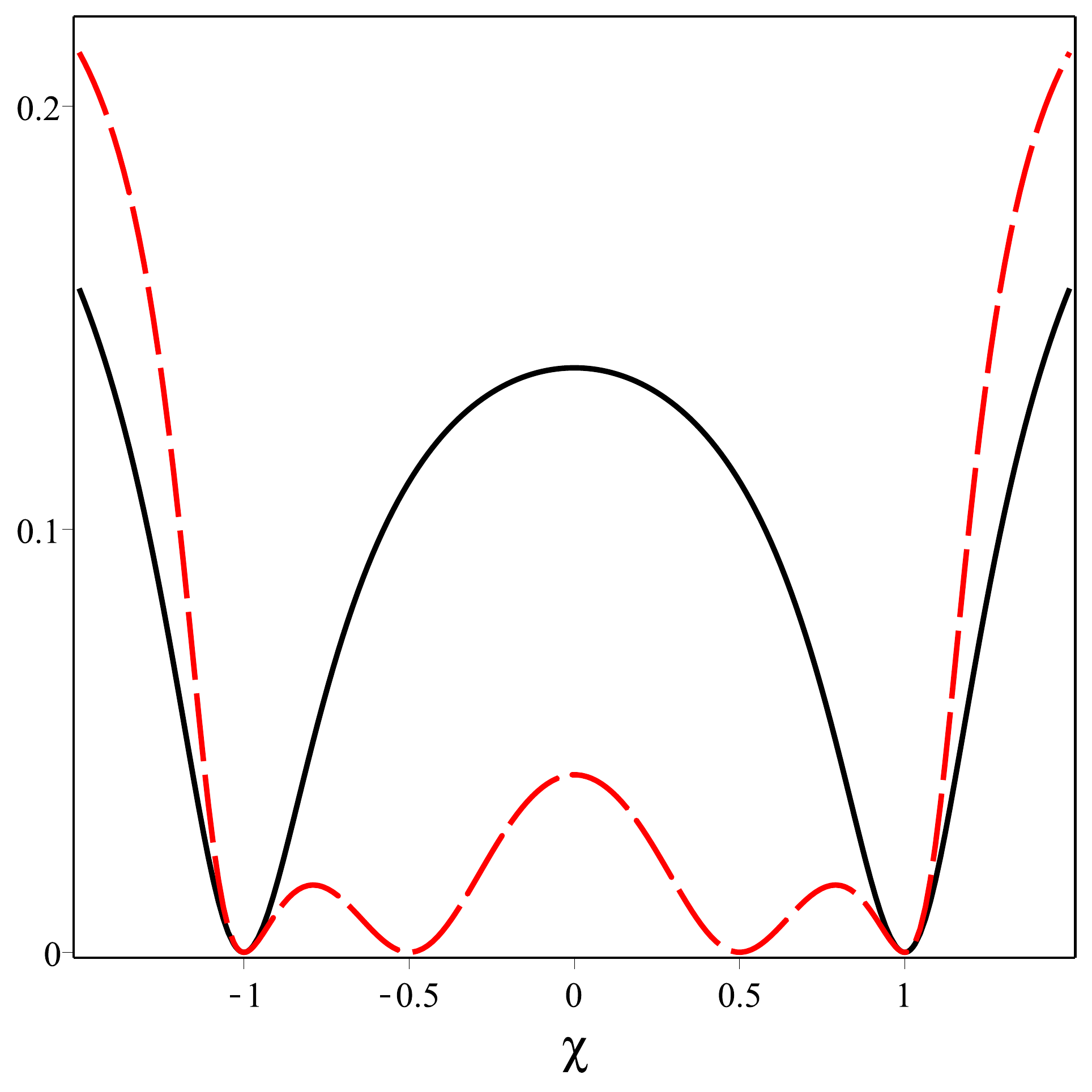}
\includegraphics[width=4.4cm,height=4.4cm]{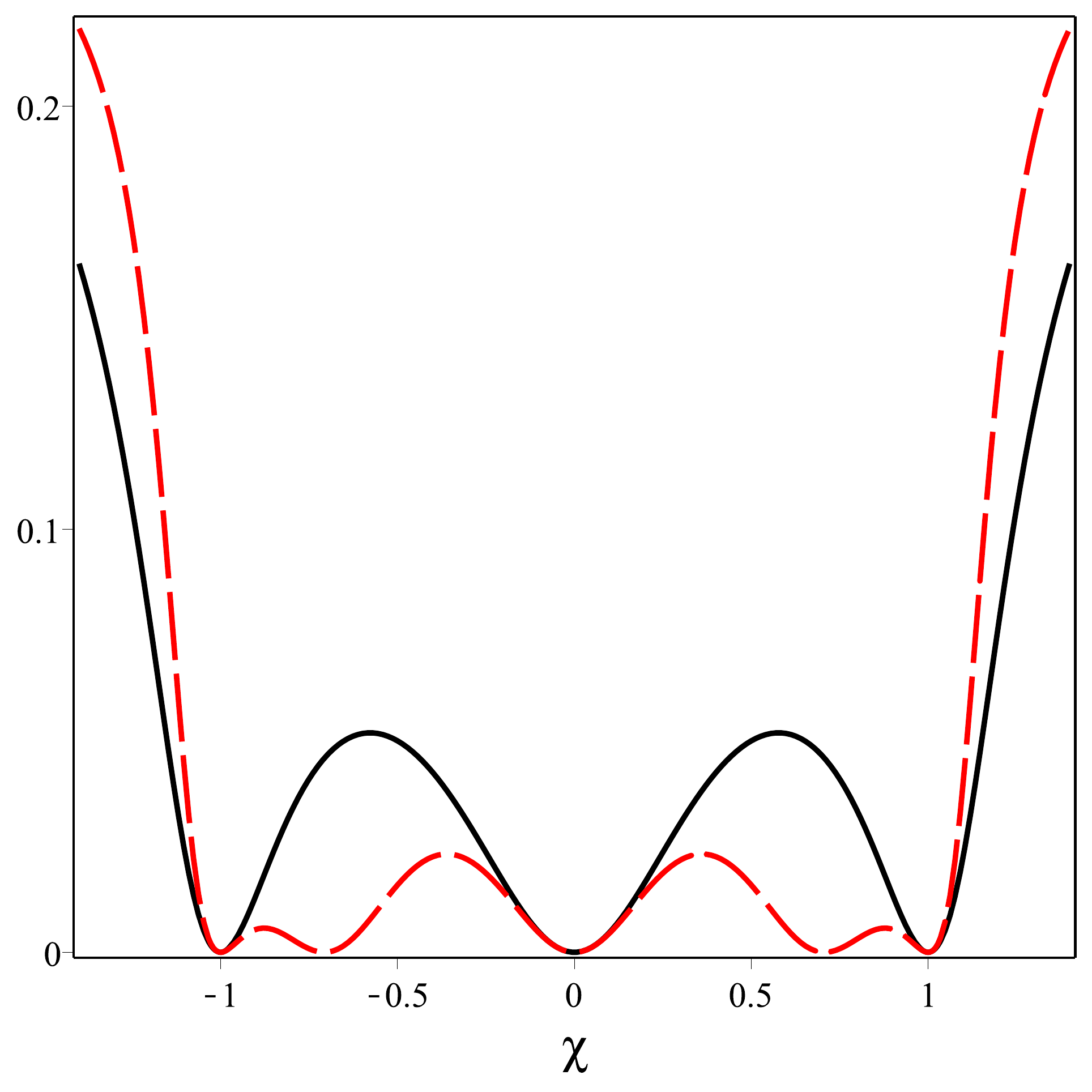}
\caption{The potential \eqref{potfp} displayed  for $a=1/2$. In the left panel one shows $\tilde{V}_{1}(\chi)$ and $\tilde{V}_{3}(\chi)$, depicted with  solid (black) and dashed (red)  lines, respectively. In the right panel one depicts $\tilde{V}_{2}(\chi)$ and $\tilde{V}_{4}(\chi)$ with  solid (black) and dashed (red)  lines, respectively.}
\label{fig:3}
\end{figure}

\begin{figure}
\includegraphics[width=4.4cm,height=4.4cm]{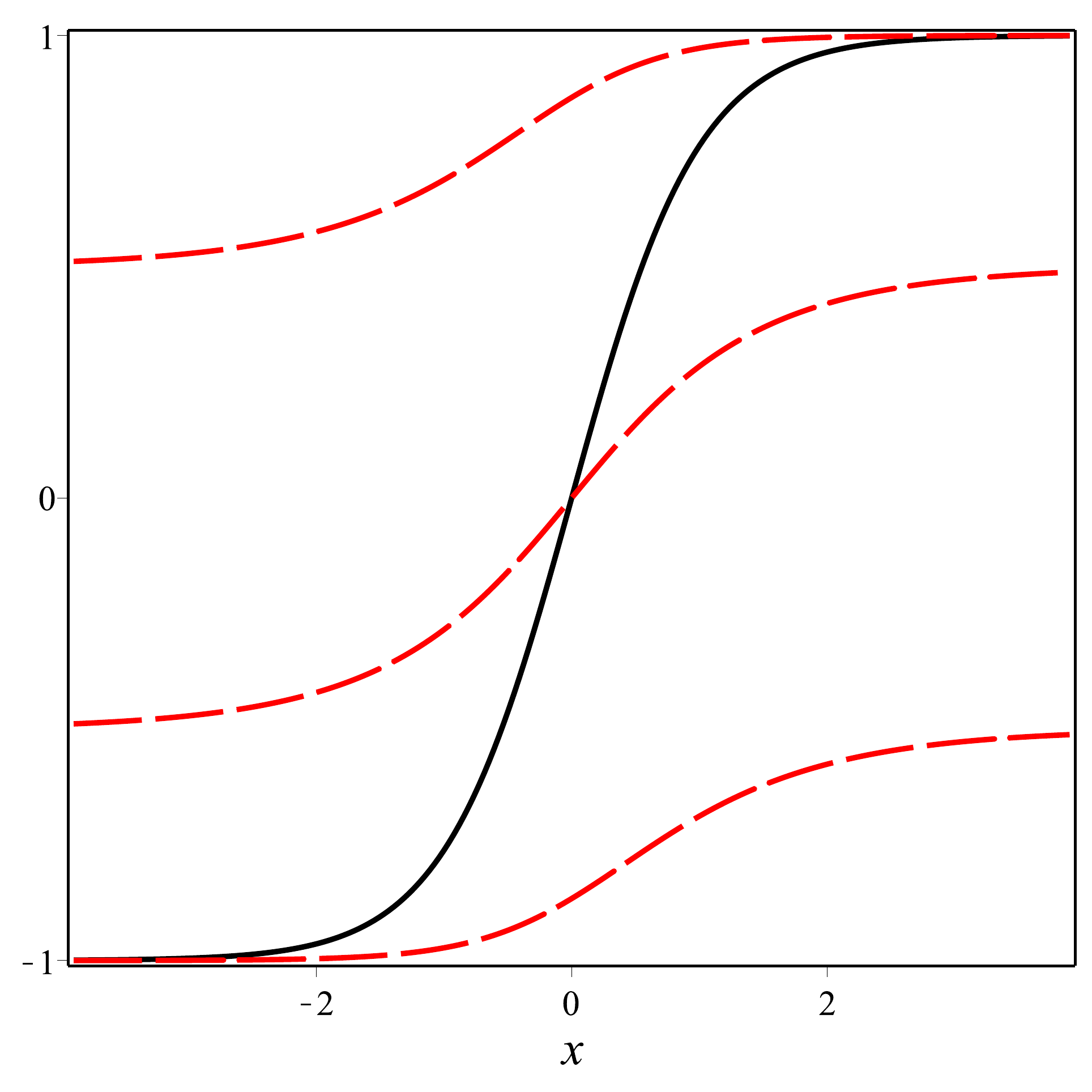}
\includegraphics[width=4.4cm,height=4.4cm]{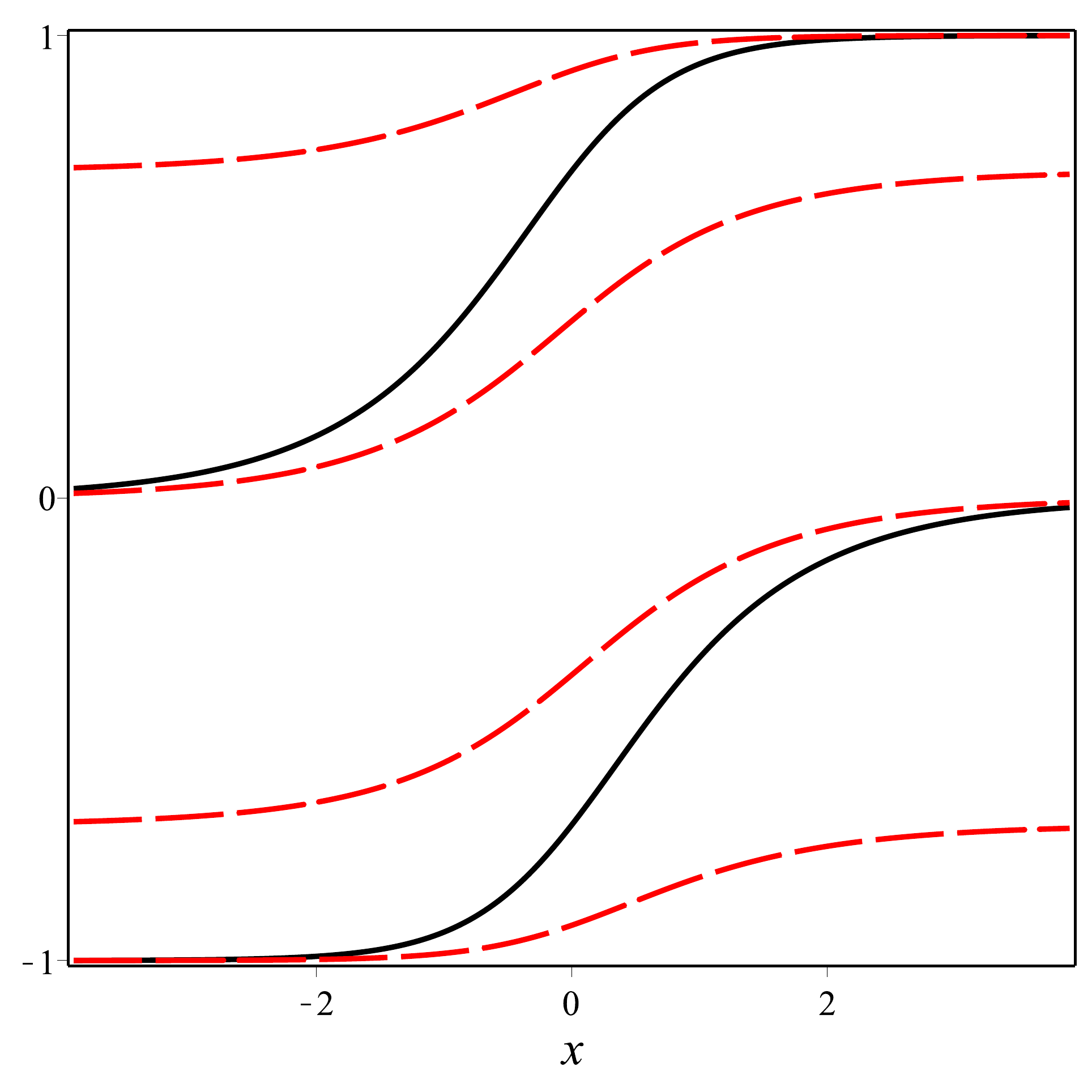}
\caption{The topological kink solutions \eqref{solfp}. In the left panel one takes $b = 1$ $(m = 0)$ depicted with solid (black), and $b = 3$ $(m = 0, 1, 2)$ represented by  dashed (red) lines. In the right panel, one assumes $b = 2$ $(m = 0, 1)$ depicted with solid (black), and $b = 4$ $(m = 0, 1, 2, 3)$ represented by dashed (red) lines.}\label{fig:4}
\end{figure}

Up to now, we have given some examples of models with polynomial interactions. Accordingly, we choose a more general deformation function which furnishes a family of polynomial potentials \cite{FPP}. That is, we consider a set of potentials involving the scalar field with higher and higher powers. In this case, for  $f_{b}(\chi)=\cos(b\arccos(\chi)-m\pi)$ we have
\be
S_{b}(\chi)=\pm \frac{1}{b}(1-\chi^2)U_{b-1}(\chi),
\ee
where $b$ takes non-null positive integer values, $m$ is a positive integer, and $U_{b-1}(\chi)$ represents the Chebyshev polynomials of second kind, given by
\be
\label{Chebyshev}
U_{b-1}(\chi)=\frac{\sin(b\, \arccos(\chi))}{\sin(\arccos(\chi))}.
\ee
The deformed potential \eqref{DP} becomes 
\be
\label{potfp}
\tilde{V}_{b}(\chi)=-a^2\left(\frac{1}{\sqrt{1+\dfrac{1}{a^2b^2}(1-\chi^2)^2 U_{b-1}^2(\chi)}}-1\right).
\ee
This determines a family of polynomial potentials controlled by the parameter $b$; and both $Models$ $I$ and $II$ described in the last two subsections, are identified as particular cases of this new set of potentials. To visualize that explicitly, one takes $b=1,2,3,4$
\ben
\tilde{V}_{1}(\chi)&=&-a^2\left(\frac{1}{\sqrt{1+\dfrac{1}{a^2}(1-\chi^2)^2}}-1\right), \nonumber\\
\tilde{V}_{2}(\chi)&=&-a^2\left(\frac{1}{\sqrt{1+\dfrac{1}{a^2}\chi^2(1-\chi^2)^2}}-1\right), \nonumber \\
\tilde{V}_{3}(\chi)&=&-a^2\left(\frac{1}{\sqrt{1+\dfrac{1}{9a^2}(1-\chi^2)^2(1-4\chi^2)^2}}-1\right),\nonumber \\
\tilde{V}_{4}(\chi)&=&-a^2\left(\frac{1}{\sqrt{1+\dfrac{1}{a^2}\chi^2(1-\chi^2)^2(1-2\chi^2)^2}}-1\right).\nonumber
\een
These potentials are described in the Fig.~\ref{fig:3},  showing two classes of models: for $b$-odd, they have a maximum at the origin, like the
$\phi^4$ model; and for $b$-even, they have a minimum at the origin, like the $\phi^6$ model. Furthermore, new topological sectors appear as $b$ increases, in such a way that the minima are given by
\be
\bar{\chi}_{b,n}=\cos\left(\frac{n\pi}{b}\right),
\ee
where $n=0,1,2..,b$.  All of static solutions are obtained by using the inverse of the deformation function
\be
\label{solfp}
\chi_{b,m}(x)=\cos\left(\frac{\arccos(\tanh(x))+m\pi}{b} \right),
\ee
where different values of $m$ furnish solutions connecting the distinct topological sectors of $\tilde{V}_{b}(\chi)$, for each specific value of $b$; for $m = 0,...,b-1$ one gets  kink solutions, and for $m = b,...,2b-1$ one gets anti-kink solutions.  In Fig.~\ref{fig:4},  we show the kink solutions corresponding to the potentials illustrated in the Fig.~\ref{fig:3}. The energy density of the solutions are given by
\be
\label{rhop}
\rho_{b,m}(x)=\frac{1}{b^2}\frac{\sin^2\left(\frac{\theta(x)+m\pi}{b} \right)\sech^2(x)}{\sqrt{1+\dfrac{1}{a^2b^2}\sin^2\left(\frac{\theta(x)+m\pi}{b} \right)\sech^2(x)}},
\ee
where we have made $\theta(x)=\arccos(\tanh(x))$, then $\theta(x)\in [0,\pi] $.

\subsection{Nonpolynomial interactions}

\subsubsection{Model III}

\begin{figure}
\includegraphics[width=4.2cm,height=4.2cm]{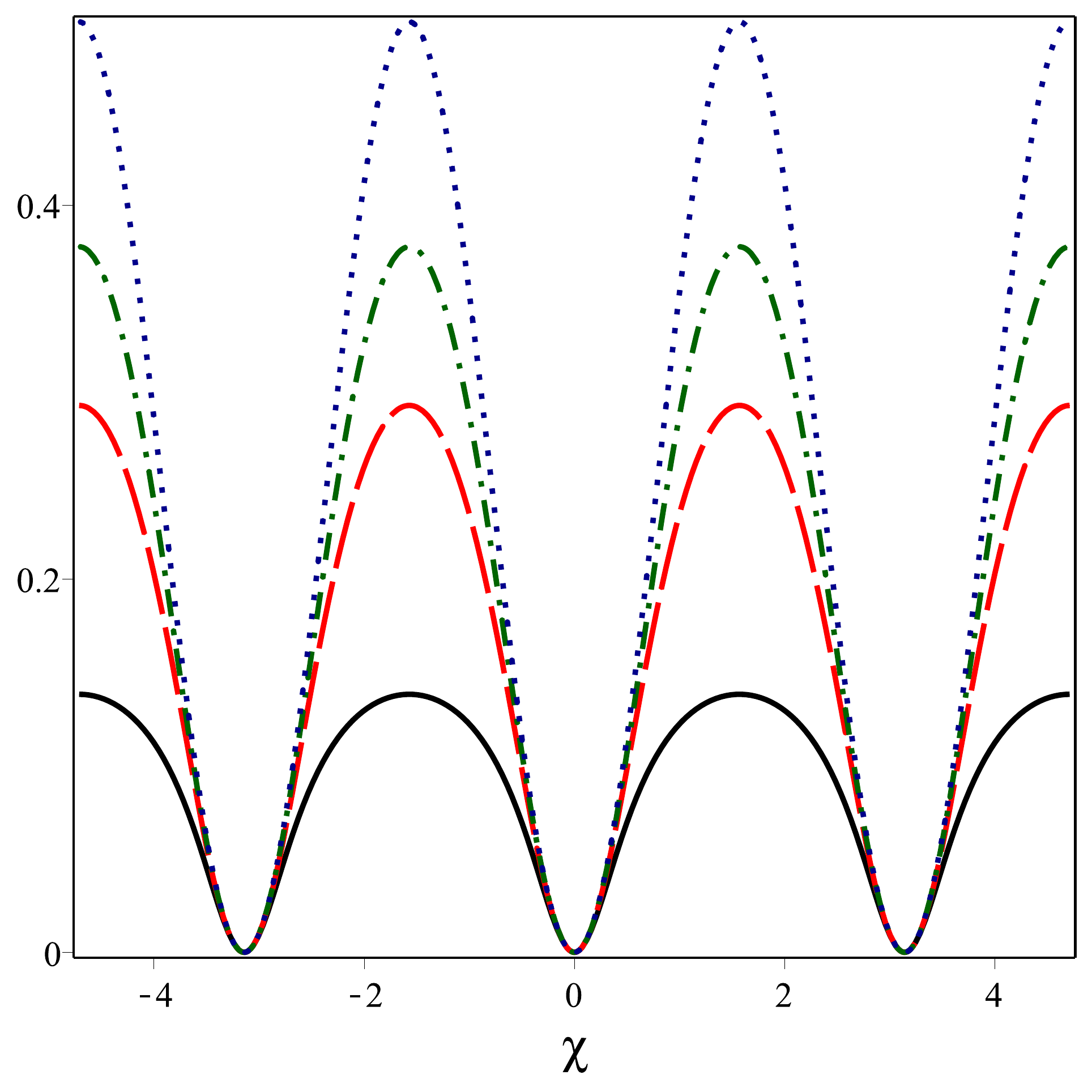}
\includegraphics[width=4.2cm,height=4.2cm]{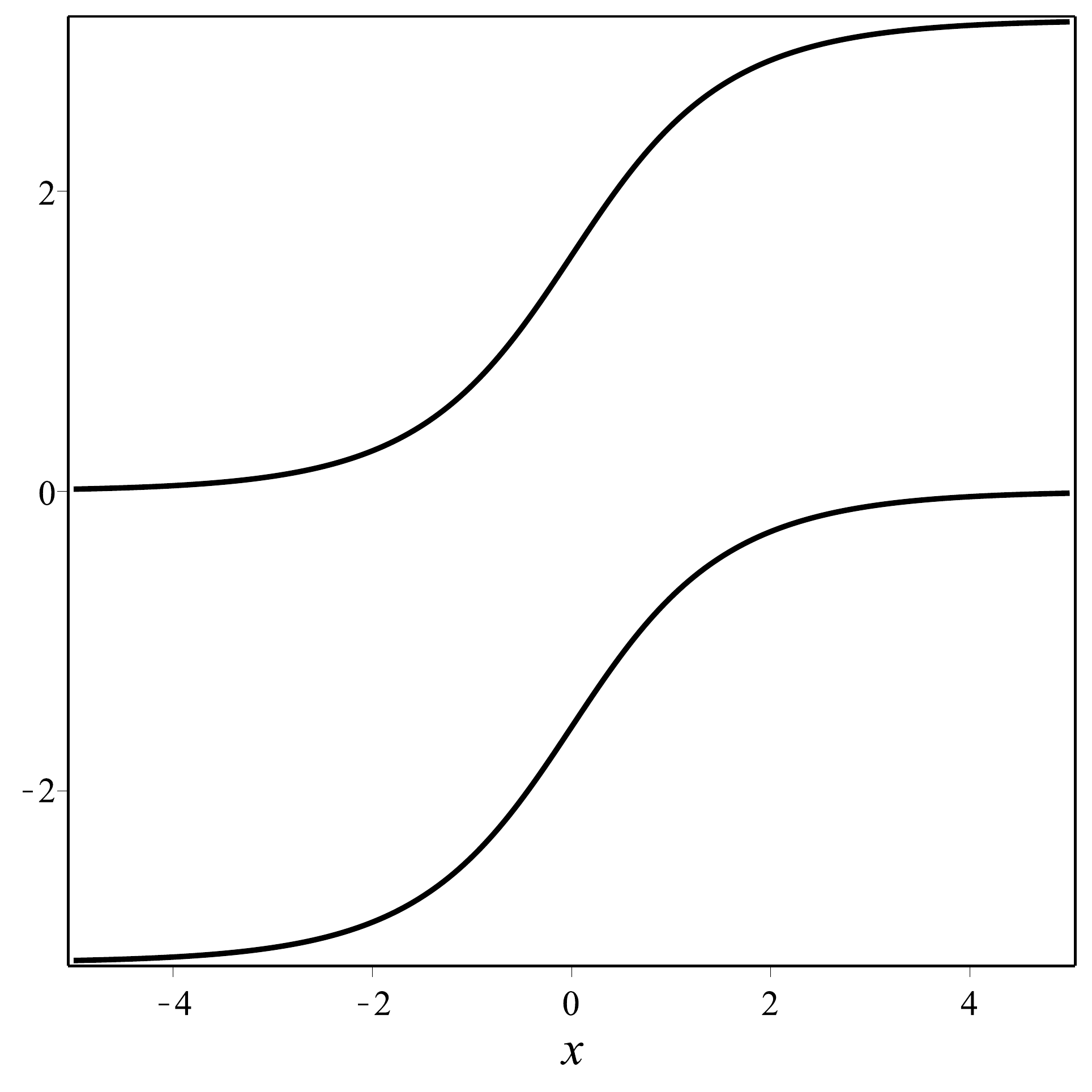}
\includegraphics[width=4.2cm,height=4.2cm]{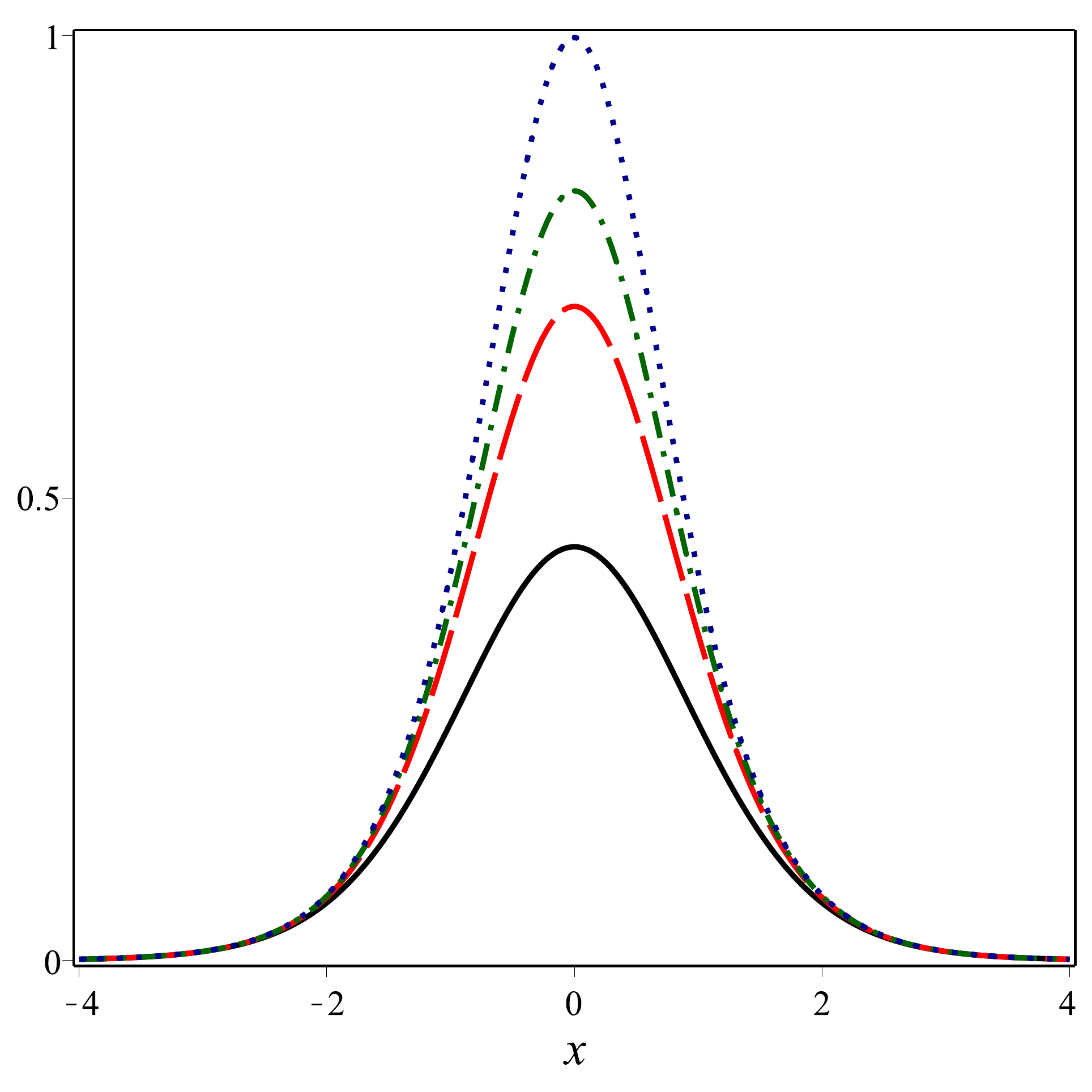}
\includegraphics[width=4.2cm,height=4.2cm]{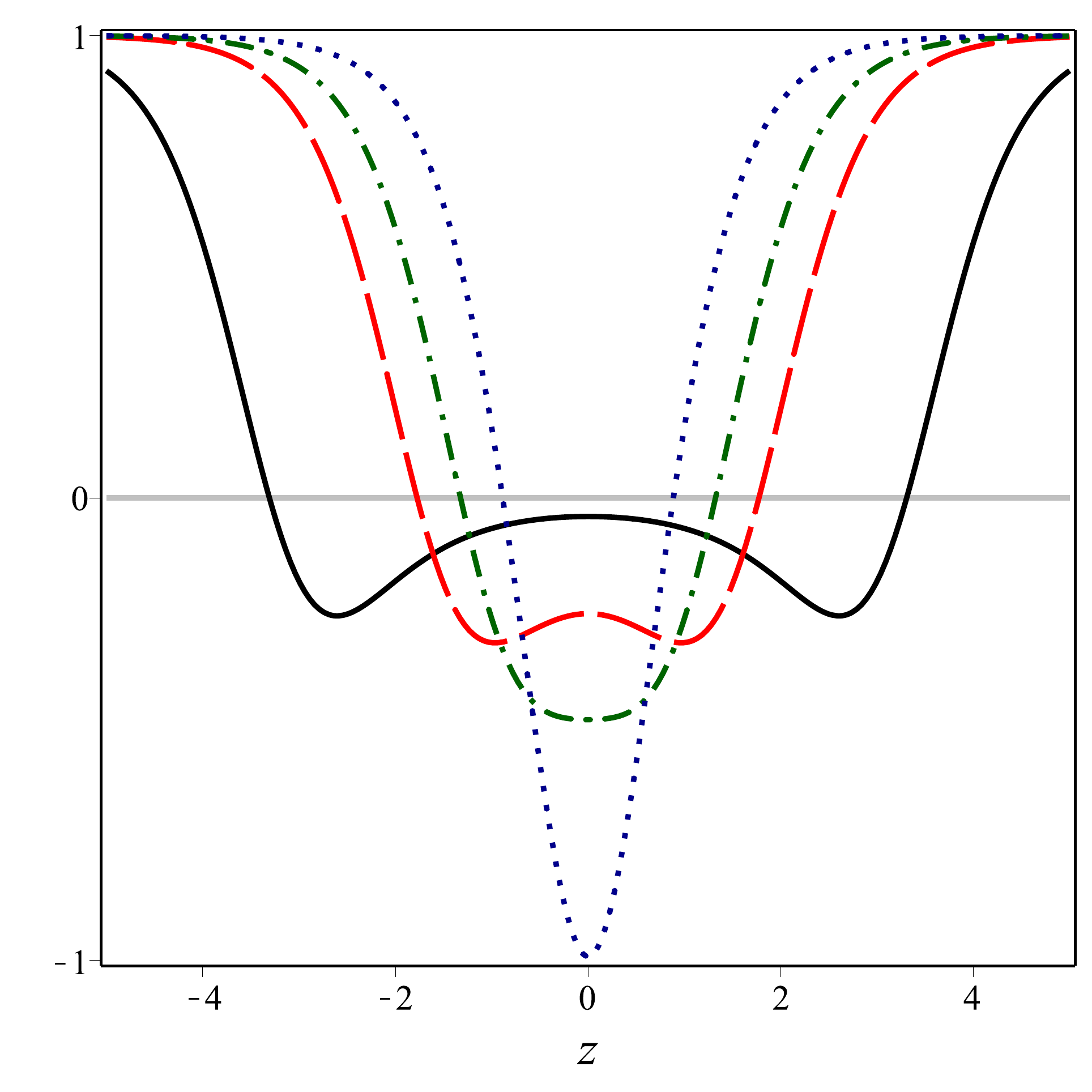}
\caption{We illustrate from left to right, the potential \eqref{potS}, the kink solutions \eqref{solS} for $k=0, 1$, the energy density \eqref{rhoS} and the stability potential \eqref{spotg}, which is evaluated numerically. We are using $a=0.5,1,1.5,15$, represented by solid (black), dashed (red), dot-dashed (green) and dotted (blue) lines, respectively.}
\label{fig:5}
\end{figure}

Let $f(\chi)=-\cos(\chi)$ be the deformation function, so
\be
S(\chi)=\sin(\chi).
\ee
The deformed potential assume the form 
\be
\label{potS}
\tilde{V}_{\sin}(\chi)=-a^2\left(\frac{1}{\sqrt{1+\dfrac{1}{a^2}\sin^2(\chi)}}-1\right),
\ee
with minima given at $\bar{\chi}=0, \pm \pi, \pm 2\pi,...$; so there is an infinity number of topological sectors which are identical and symmetrical; their solutions are 
\be
\label{solS}
\chi_{k}(x)=\mp \arccos(\tanh(x))+k\pi, 
\ee
where $k=0,\pm 1,\pm 2$,..., each value of $k$ provides solutions of one specific sector of $\tilde{V}_{\sin}$; and the energy density is
\be
\label{rhoS}
\rho(x)=\frac{\sech^2(x)}{\sqrt{1+\dfrac{1}{a^2}\sech^2(x)}}.
\ee
So by taking into account the fact that all topological sectors are identical; they have the same energy density and stability potential, in which there is  a change from a maximum to a minimum at the origin, when $a^2=2$. These quantities are plotted in the Fig.~\ref{fig:5},  for different values of $a$. Note that,  $1/a^2 << 1$ implies  results coinciding with a sine-Gordon model \cite{r1}, that is,  $V(\chi)\approx \frac{1}{2}\sin(\chi)^2$. The energy density becomes  $\rho(x) \approx \sech(x)^2$; and the study of stability leads to the modified  P\"oschl-Teller potential,
\be
U(x) \approx 1-2\,\sech^2(x),
\ee
which has only one bound state, the zero mode.

\subsubsection{Model IV}

\begin{figure}
\includegraphics[width=6.5cm,height=5cm]{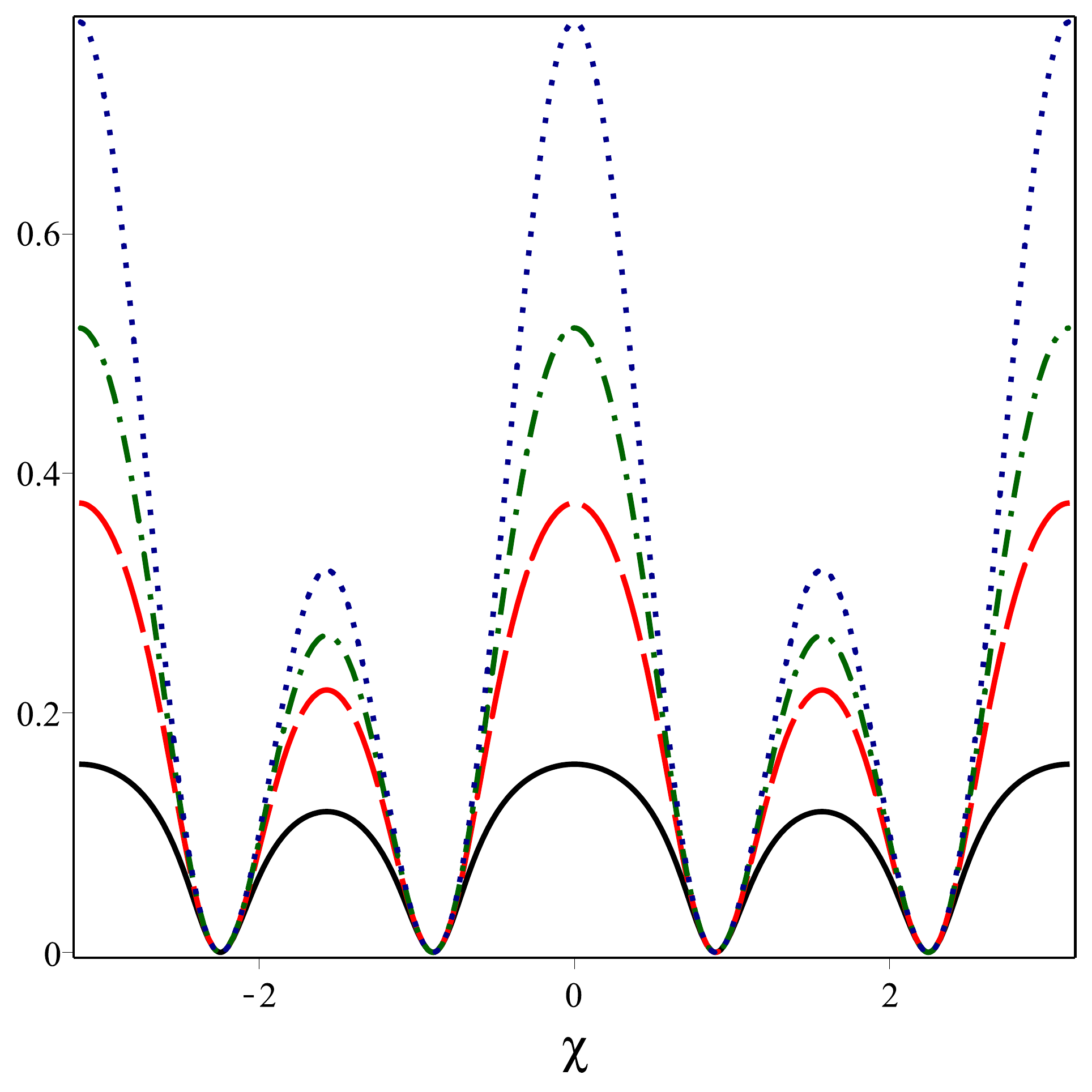}
\includegraphics[width=6.5cm,height=5cm]{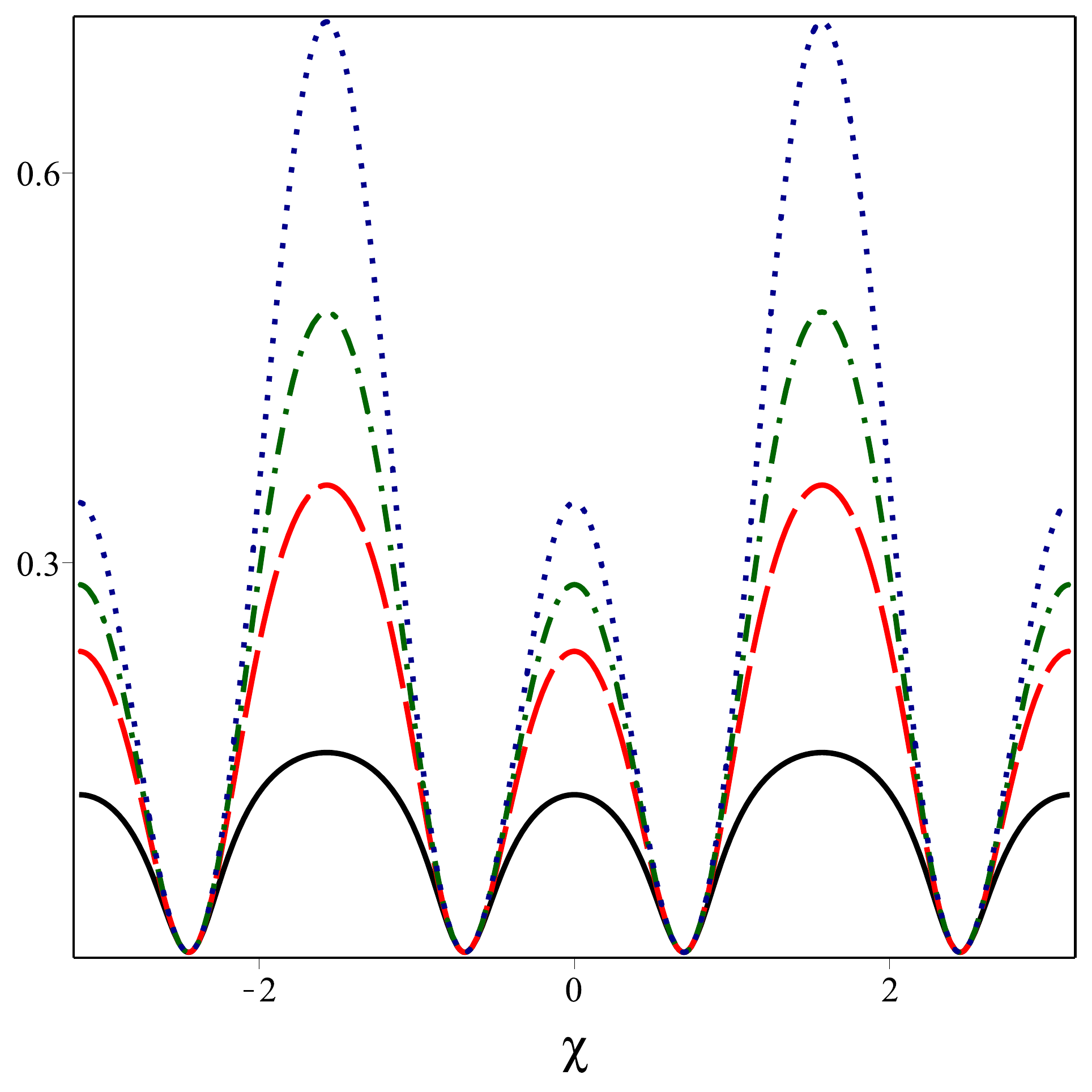}
\caption{Potential $\tilde{V}_{r}(\chi)$, Eq.~\eqref{potDS}, displayed  for $a=0.5,1,1.5,15$, represented by solid (black), dashed (red), dot-dashed (green) and dotted (blue) lines, respectively. One makes $r=0.8$, in the top panel, while $r=1.2$ in the bottom panel.}\label{fig:6}
\end{figure}

Next, if we choose the following functions $f_{r}(\chi)=r\tan(\chi)$ and $1/f_{r}(\chi)=\frac{1}{r}\cot(\chi)$, with positive real $r$, we get the same expression for $S(\chi)$,
\be
S_{r}(\chi)=\frac{1}{r}\left((1+r^2)\cos^2(\chi)-r^2\right),
\ee
and obviously the same deformed potential
\be
\label{potDS}
\tilde{V}_{r}(\chi)=-a^2\left(\frac{1}{\sqrt{1+\dfrac{1}{a^2r^2}\left((1+r^2)\cos^2(\chi)-r^2\right)^2}}-1\right)\!.
\ee
As can be seen in Fig.~\ref{fig:6}, this model has aspect fairly similar to the double sine-Gordon \cite{DSG}, which got an infinity number of two distinct sectors, one large and other small, whose behavior depends on $r$ and $a$. This property basically defines the difference between both double sine-Gordon and sine-Gordon models. For the region $r\in (0,1)$ the central sector of $\tilde{V}_{r}(\chi)$ is a large one, but it becomes a small one for $r\in (1, \infty)$. At $r=1$, both sectors become equal, so the system becomes of the sine-Gordon type. Moreover, this potential has minima given by
\be
\bar{\chi}_{r,k}=\pm \arctan(1/r)+k\pi, \,\,\,\, k=0,\pm 1, \pm 2,...
\ee
The two distinct topological sectors are characterized by two distinct solutions (small and large), one is  obtained by the deformation function $f_{r}(\chi)$ and  the other by $1/f_{r}(\chi)$ . They are
\ben
\label{solDS1}
\chi_{r,k}^{(1)}(x)&=&\pm \arctan\left(\frac{1}{r}\tanh(x)\right)+k\pi, \\ 
\label{solDS2}
\chi_{r,k}^{(2)}(x)&=&\mp \arccot(r \,\tanh(x))+k\pi.
\een
The energy densities for the two distinct categories of solutions above are given respectively by
\ben
\label{rhoDS1}
\rho_{r}^{(1)}(x)&=&\dfrac{r^2\sech^4(x)}{ \left(r^2+\tanh^2(x)\right)^2\sqrt{1+\dfrac{r^2\sech^4(x)}{a^2\left(r^2+\tanh^2(x)\right)^2}}}, \nonumber \\ \\
\label{rhoDS2}
\rho_{r}^{(2)}(x)&=&\dfrac{r^2\sech^4(x)}{ \left(1+r^2\tanh^2(x)\right)^2\sqrt{1+\dfrac{r^2\sech^4(x)}{a^2\left(1+r^2\tanh^2(x)\right)^2}}}. \nonumber \\
\een
Observe that, for $r=1$ one gets: $\rho_{r}^{(1)}(x)= \rho_{r}^{(2)}(x)$. One also notes that for $1/a^2<<1$, one obtains the double sine-Gordon  studied in Ref.~\cite{NFSG}.

\subsubsection{Family II}

\begin{figure}
\includegraphics[width=6.5cm,height=5cm]{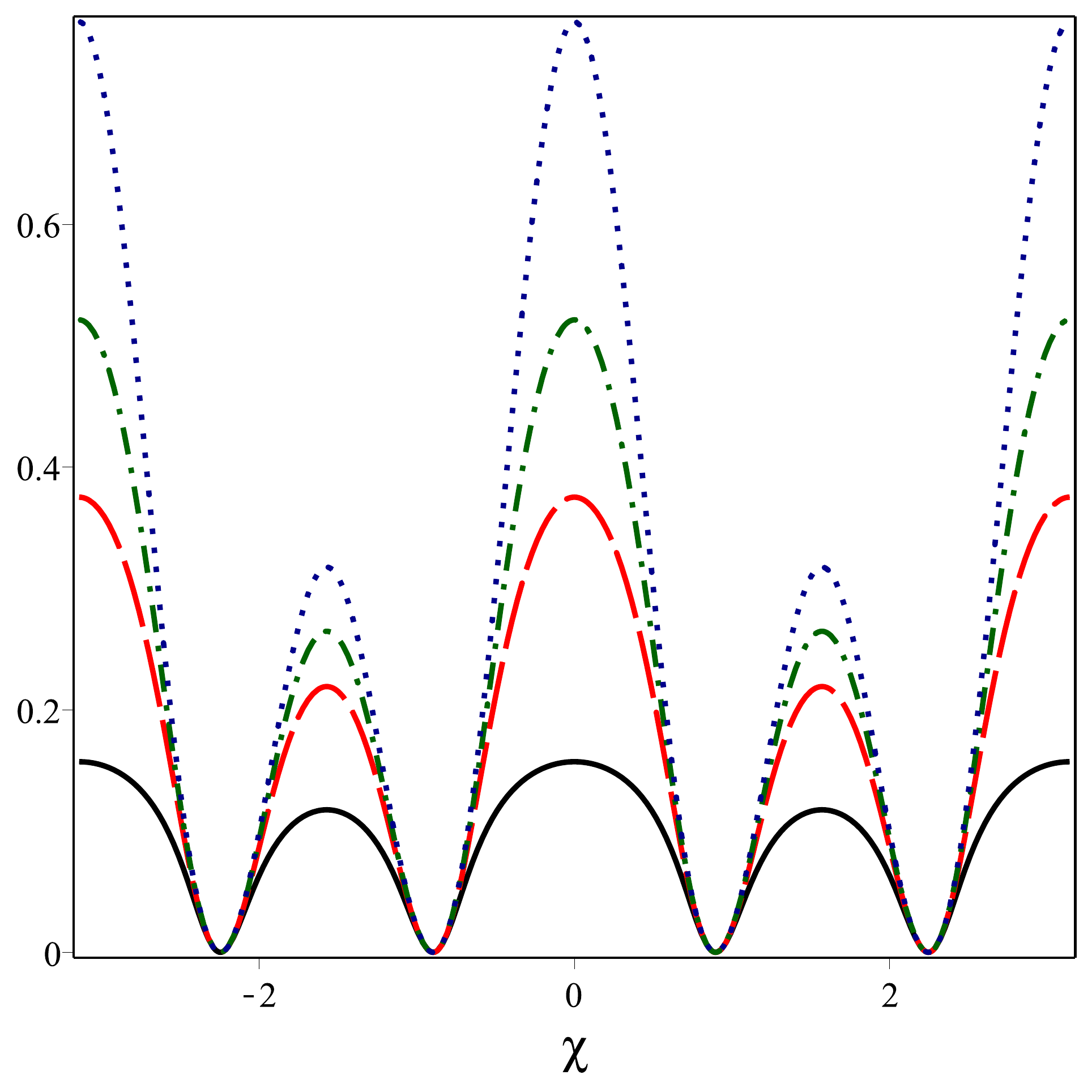}
\includegraphics[width=6.5cm,height=5cm]{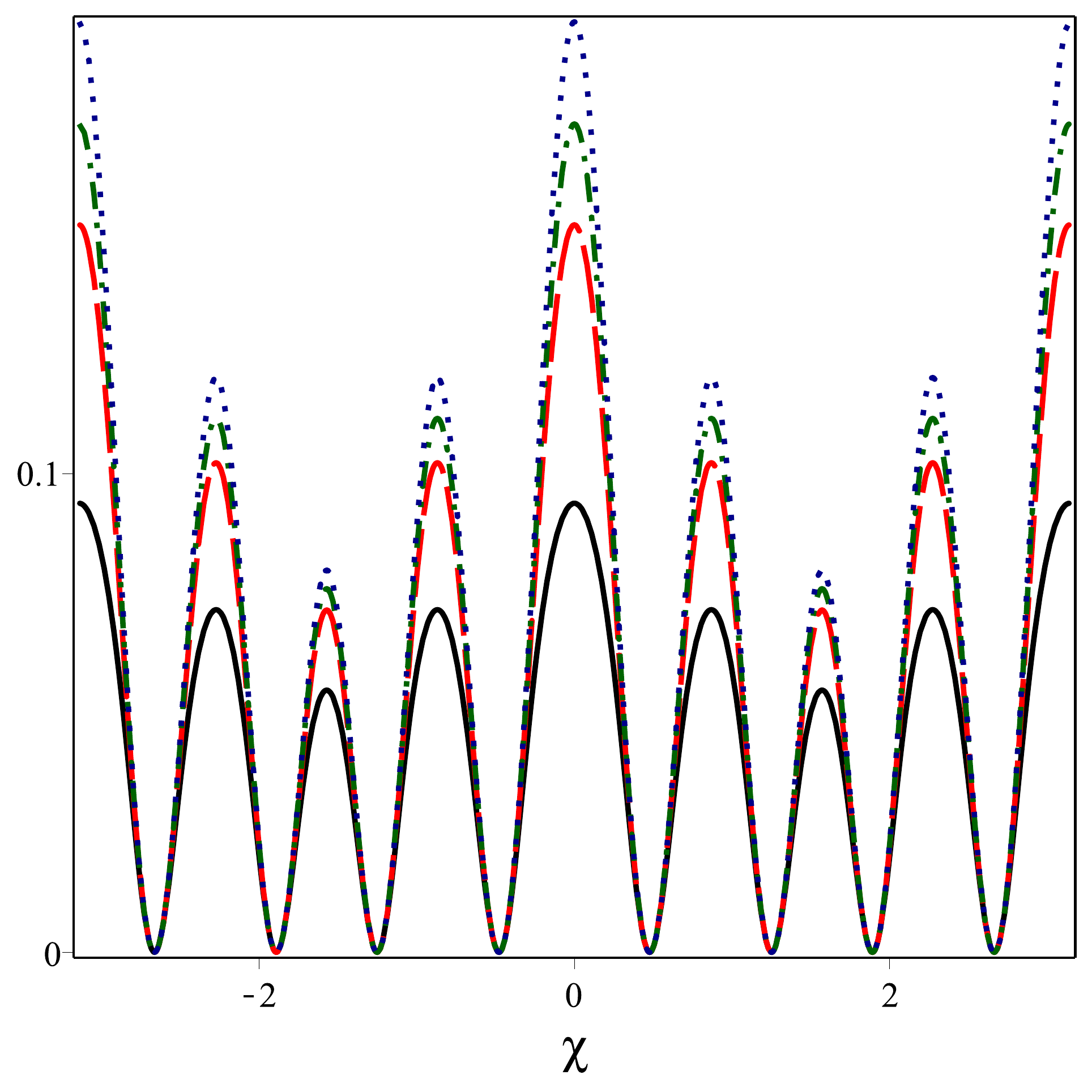}
\caption{Potential $\tilde{V}_{b,r}(\chi)$, Eq.~\eqref{potFS}, displayed  for $r=0.8$ and $a=0.5,1,1.5,15$, represented by solid (black), dashed (red), dot-dashed (green) and dotted (blue) lines, respectively. In the top panel, one makes $b=1$ and in the bottom panel, $b=2$.}\label{fig:7}
\end{figure}

Let us now propose other deformation functions. For instance, 
\ben
f_{b,r}(\chi)&=&\tan(b\arctan(r\tan(\chi))), \\
\frac{1}{f_{b,r}(\chi)}&=&\cot(b\arctan(r\tan(\chi))), \\
g_{b,r}(\chi)&=&\tan\left(b\arctan\left(\frac{1}{r}\cot(\chi)\right)\right), \\
\frac{1}{g_{b,r}(\chi)}&=&\cot\left(b\arctan\left(\frac{1}{r}\cot(\chi)\right)\right).
\een
They lead to 
\ben
\label{Sbr}
S_{b,r}(\chi)=\pm \frac{1}{br}\left((1-r^2) \cos^2(\chi)+r^2\right)\times \nonumber \\ 
\left(2 \cos(b\arctan(r\tan(\chi)))^2-1\right),
\een
where  $b=1,2,...,$ and $r$ is positive and real; the bottom sign $(-)$ of Eq.~\eqref{Sbr} is generated by the functions $g_{b,r}(\chi)$ and $1/g_{b,r}(\chi)$, with $b$-even. The deformed potential gets the form
\be
\label{potFS}
\tilde{V}_{b,r}(\chi)=-a^2\left(\frac{1}{\sqrt{1+\dfrac{1}{a^2}S_{b,r}(\chi)^2}}-1\right),
\ee
with minima at
\be
\bar{\chi}_{b,r,m,k}=\arctan\left(\frac{1}{r}\tan\left((1+2m)\frac{\pi}{4b}\right)\right)+k\pi,
\ee
where $m=0,...,2b-1$, and $k=0,\pm 1,\pm 2,...$ This model is displayed in Fig.~\ref{fig:7}, showing the
arising of new topological sectors as $b$ increases, that leads to a family of sine-Gordon-like potentials described by $(b + 1)$ different sectors which repeating themselves infinitely: for $b = 1$ one gets to a model similar to the double sine-Gordon, which contains two distinct topological sectors; for $b = 2$ one gets to a model similar to the triple sine-Gordon, which contains three distinct topological sectors; for $b = 3$ one gets to a model similar to the quadruple sine-Gordon, which contains four distinct topological sectors; and so forth. The parameter $r$ performs a similar role as before in subsection {\it Model IV}, and it controls the position of the minima and the width of the topological sectors. We also note that all the sectors become equal for $r=1$.

Once there is four different deforming functions driving to the same potential, then there is four distinct kinds of topological solutions, given respectively by
\ben
\label{solFS1}
\chi_{b,r,l,k}^{(1)}(x)&=&\pm \arctan \left(\frac{1}{r}\tan\left(F_{b,l}(x)\right)\right)+k\pi \\
\label{solFS2}
\chi_{b,r,l,k}^{(2)}(x)&=&\mp \arctan \left(\frac{1}{r}\tan\left(G_{b,l}(x)\right)\right)+k\pi \\
\label{solFS3}
\chi_{b,r,l,k}^{(3)}(x)&=& \mp \arccot \left(r\tan\left(F_{b,l}(x)\right)\right)+k\pi \\
\label{solFS4}
\chi_{b,r,l,k}^{(4)}(x)&=& \pm \arccot \left(r\tan\left(G_{b,l}(x)\right)\right)+k\pi
\een
with
\ben
F_{b,l}(x)&=& \frac{1}{b}\arctan(\tanh(x))+\frac{l\pi}{b}, \nonumber\\
G_{b,l}(x)&=& \frac{1}{b}\arccot(\tanh(x))+\frac{l\pi}{b}. \nonumber
\een
where $l,k=0,\pm 1, \pm 2,...$  allow the solutions to appear in distinct sectors of $\tilde{V}_{b,r}(\chi)$, with fixed $b$ and $r$.
 
The  corresponding energy densities are given by
\be
\rho_{b,r,l,k}^{(i)}(x)=\frac{\left(S_{b,r,l,k}^{(i)}(x)\right)^2}{\sqrt{1+\dfrac{1}{a^2}\left(S_{b,r,l,k}^{(i)}(x)\right)^2}},
\ee
where $i=1,2,3,4$ refers to the four distinct types of solutions above, and the functions $S_{b,r,l,k}^{(i)}(x)$ are
\ben
\label{SFS}
S_{b,r,l,k}^{(1)}(x)&=&\frac{1}{br}\frac{\sech^2(x)\sec^2(F_{b,l})}{\left(1+\tanh^2(x)\right)\left(1+\dfrac{1}{r^2}\tan^2(F_{b,l})\right)}, \nonumber \\
S_{b,r,l,k}^{(2)}(x)&=&-\frac{1}{br}\frac{\sech^2(x)\sec^2(G_{b,l})}{\left(1+\tanh^2(x)\right)\left(1+\dfrac{1}{r^2}\tan^2(G_{b,l})\right)}, \nonumber \\
S_{b,r,l,k}^{(3)}(x)&=&-\frac{r}{b}\frac{\sech^2(x)\sec^2(F_{b,l})}{\left(1+\tanh^2(x)\right)\left(1+r^2\tan^2(F_{b,l})\right)}, \nonumber\\
S_{b,r,l,k}^{(4)}(x)&=&\frac{r}{b}\frac{\sech^2(x)\sec^2(G_{b,l})}{\left(1+\tanh^2(x)\right)\left(1+r^2\tan^2(G_{b,l})\right)}. \nonumber
\een

\subsubsection{Another model}

We can also consider fields that interact hyperbolically, as recently investigated in Ref.~\cite{lima}. We exemplify this case with the model
\be
W_{\phi}=\frac{1-\sinh^2(\phi)}{\sqrt{1+\dfrac{1}{a^2}\left(1-\sinh^2(\phi)\right)^2}}.
\ee
We then use \eqref{potw} to get
\be
\label{poth}
V(\phi)=-a^2\left(\frac{1}{\sqrt{1+\dfrac{1}{a^2}\left(1-\sinh^2(\phi)\right)^2}}-1\right).
\ee
This potential has two minima at $\bar{\phi}=\pm {\rm {arcsinh}}(1)$ and one maximum at the origin, independently of $a$. The first-order equation \eqref{fow} leads to
\be
\phi'=1-\sinh^2(\phi),
\ee
and has the solution 
\be
\label{solh}
\phi(x)= {\rm arctanh}\left(\frac{1}{\sqrt{2}}\tanh\left(\sqrt{2}x\right)\right).
\ee
The energy density \eqref{rhofo} gives
\be
\label{rhh}
\rho(x)=\frac{4}{ \left(1+\cosh^2\left(\sqrt{2}x\right)\right)^2\!\!\sqrt{1+\dfrac{4}{a^2\left(1+\cosh^2\left(\sqrt{2}x\right)\right)^2}}}.
\ee

\begin{figure}
\includegraphics[width=4.2cm,height=4.2cm]{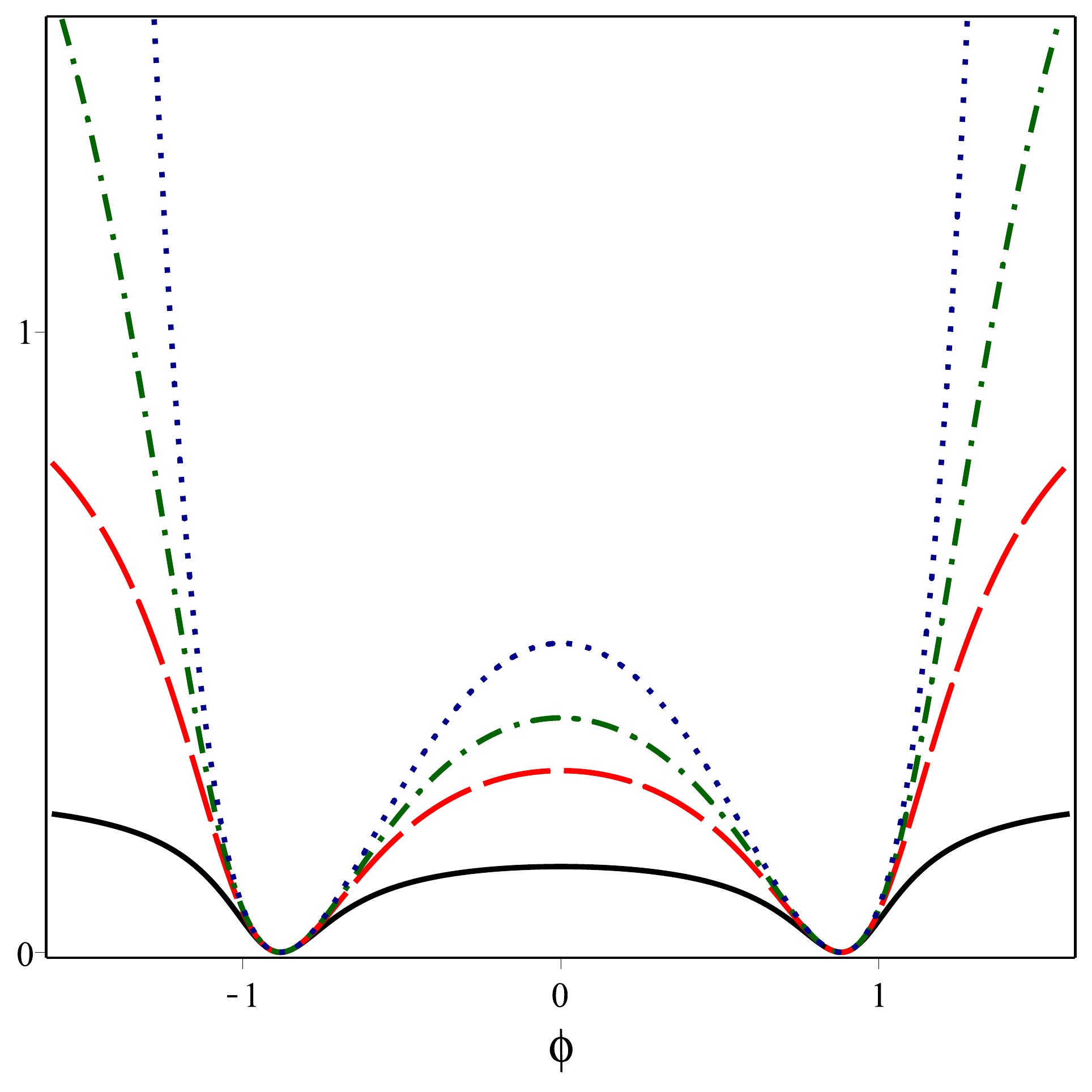}
\includegraphics[width=4.2cm,height=4.2cm]{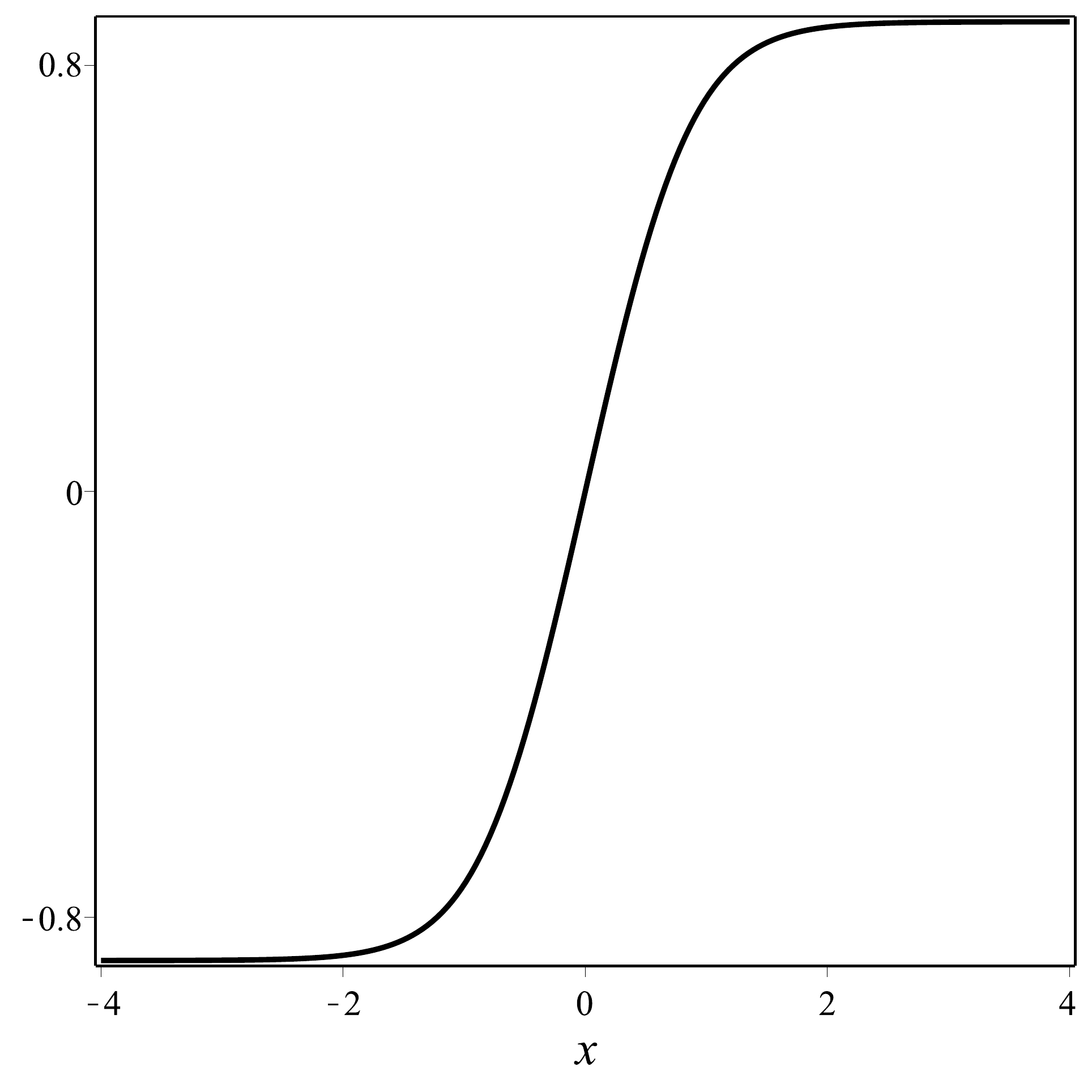}
\includegraphics[width=4.3cm,height=4.2cm]{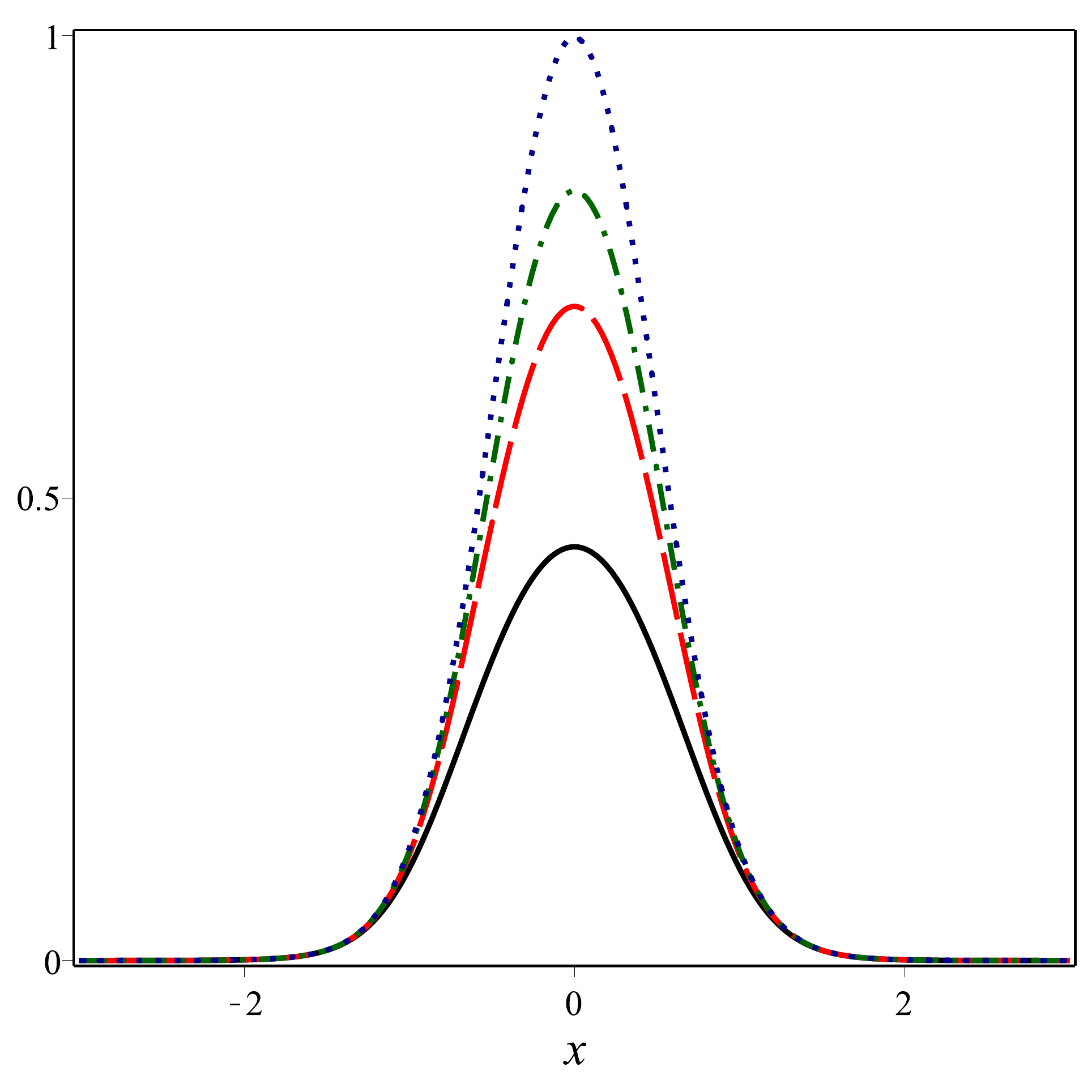}
\includegraphics[width=4.2cm,height=4.2cm]{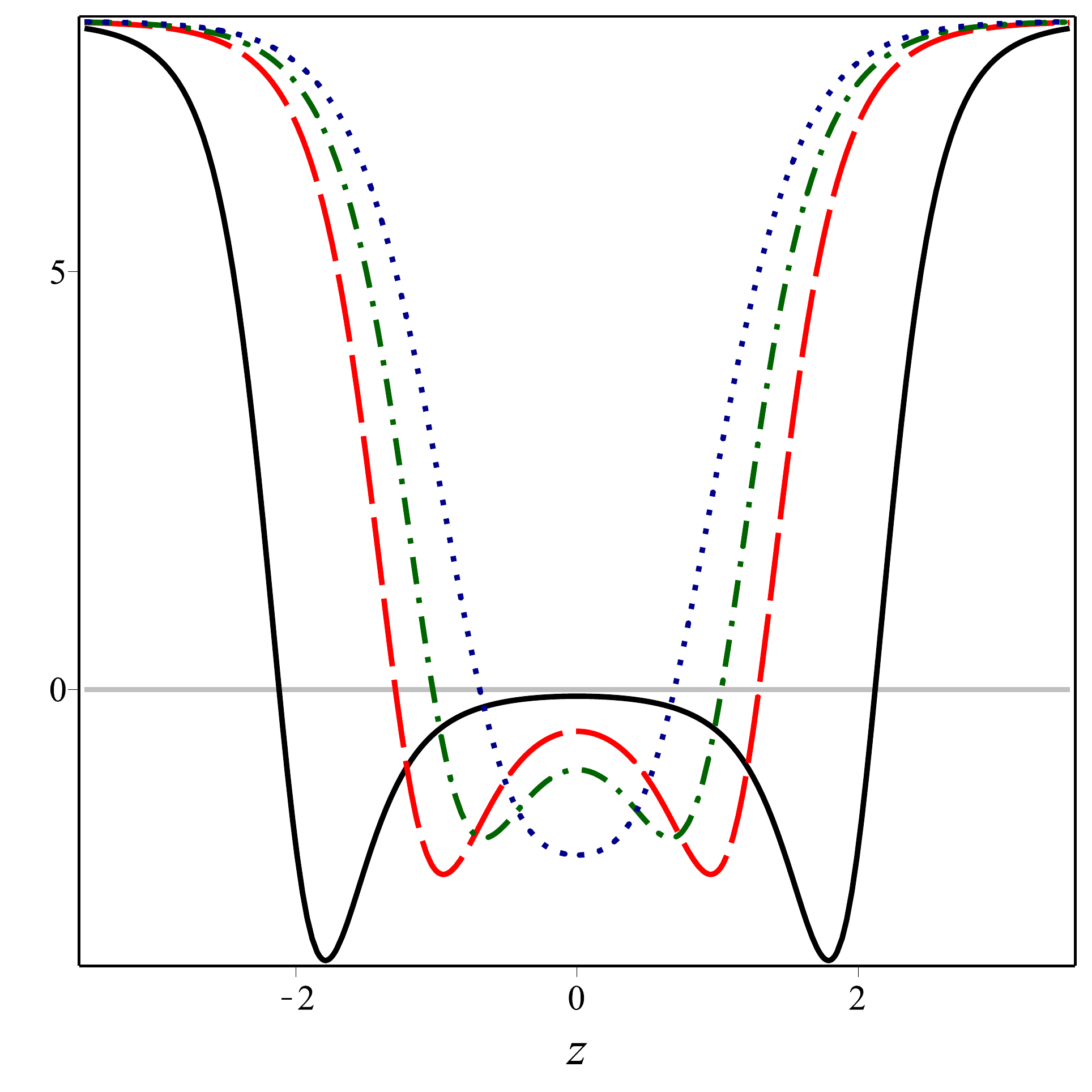}
\caption{WE illustrate from left to right, the potential \eqref{poth}, the kink solution \eqref{solh}, the energy density \eqref{rhh} and the stability potential \eqref{spotg}, which is evaluated numerically. We are using $a=0.5,1,1.5,15$, represented by solid (black), dashed (red), dot-dashed (green) and dotted (blue) lines, respectively.}
\label{fig:8}
\end{figure}

In Fig.~\ref{fig:8} we illustrate the potential \eqref{poth}, the solution \eqref{solh}, the energy density \eqref{rhh}, and the stability potential calculated numerically through \eqref{spotg}, for several values of $a$. Furthermore, if $1/a^2 << 1$,  these quantities give similar results to the ones obtained by the standard theory with potential $V(\phi)= \frac{1}{2}\left(1-\sinh(\phi)^2\right)^2$; for instance, 
\be
\rho(x)=\frac{4}{ \left(1+\cosh^2\left(\sqrt{2}x\right)\right)^2},
\ee
and
\be 
U(x)= \frac{4\left(2\cosh^4\left(\sqrt{2}x\right)-5\cosh^2\left(\sqrt{2}x\right)+1\right)}{ \left(1+\cosh^2\left(\sqrt{2}x\right)\right)^2}.\;\;
\ee

This example shows that we can consider other models having hyperbolic interactions, in a way similar to the cases considered above, for polynomial and trigonometric interactions. 

\section{Comments and conclusions}\label{sec-com}

In this paper, we investigated two-dimensional scalar field models which develop generalized dynamics of the Dirac-Born-Infeld type. We introduced a first-order treatment that helps us to solve the equation of motion and show that the solutions  are linearly stables. 

In order to illustrate the procedure, we constructed two distinct kinds of potentials, one describing polynomial interactions and the other nonpolynomial ones, obtained by directly using the deformation procedure adopted for systems with noncanonical dynamics \cite{DMGD}. Among the polynomial interactions we highlighted the $\phi^4$-like and $\phi^6$-like models, and among the nonpolynomial interactions we highlighted the sine-Gordon, the double sine-Gordon and another model, described by the hyperbolic sine. Although the several models here introduced have nonlinear kinetic term and noncanonical potential, controlled by a parameter $a$, they have kinklike solutions equivalent to ones obtained by models with standard kinetic term and canonical potential, independently of $a$. However, their energy densities and stability potentials are modified, being $a$-dependent. We also noticed that in the limit where $a$ is very large, most of the outcome converge to those already established in the literature.

The results of the work are of current interest, since modifications on the standard dynamics for the DBI one are widely applicable in physics, as we see, for instance, in Refs.~\cite{Brown,Sarangi,Babichev2,VortexBI,Andrews,Garcia}. As future perspective for this work, it seems of interest to extend the current investigation to include supersymmetry. Since the results are obtained from a first order formalism, it seems that we can include fermions to work with supersymmetric models, enlarging the scope of the present investigation. We can also apply this study to investigate analytical compact solutions, since such structures have attracted attention in different contexts in
Refs.~ \cite{Rosenau,Adam,DMGD,compA,compN}.

Because the kinklike solutions found above are similar to the solutions that appear in standard models, one thinks that kink-antikink collisions can be implemented in a similar way, along the lines of Refs.~\cite{scater}. This issue may perhaps show how the DBI dynamics contribute to change the scenario one gets in the case of collisions under standard dynamics.

Another line of investigation concerns the fact that General Relativity faces problems near the Planck scale,
and this indicates the need of modifications at high energies. In this sense, DBI inspired modifications of gravity may be used to regularize the gravitational dynamics, inducing the presence of regular black holes spacetimes. This possibility has been recently reviewed in \cite{olmo} and we hope that the current study may inspire new investigations in the subject. In a similar way, we can also add the scalar field models here investigated in a five-dimensional warped geometry as source field models to control braneworld scenarios with a single extra dimension of infinite extent, following the lines of Refs.~\cite{Randall,brane,thickbrane,split,BO}. In the cosmological context, the current models may be used to investigate dark energy \cite{de}, dark matter \cite{dm} and interactions between them, if one extends the models to the case of two or more fields. We hope to further report on this in the near future.

\section*{Acknowledgments}

This work is partially supported by CNPq, Brazil. DB acknowledges support from grants CNPq:455931/2014-3 and CNPq:306614/2014-6, EEML acknowledges support from grant CNPq:160019/2013-3, and LL acknowledges support from grants CNPq:307111/2013-0 and CNPq:447643/2014-2.


\end{document}